\documentclass[useAMS,usenatbib]{mn2e}

\usepackage{graphicx}	
\usepackage{amsmath}	
\usepackage{amssymb}	
\usepackage{multicol}        
\usepackage{bm}		
\usepackage{pdflscape}	
\usepackage{xcolor}

\newcommand\degree{^{\circ}}

\usepackage[T1]{fontenc}
\usepackage{ae,aecompl}

\usepackage{newtxtext,newtxmath}

\date{Last updated 2020 June 10; in original form 2013 September 5}

\pubyear{2022}

\title[Wind bubbles as cosmic-rays re-accelerators]{
Stellar wind bubbles of OB stars as Galactic cosmic-ray re-accelerators
}

\author[D. M.-A.~Meyer]
       {D. M.-A.~Meyer \\ 
       Institute of Space Sciences (ICE, CSIC), 
       Campus UAB, Carrer de Can Magrans s/n, 
       08193 Barcelona, Spain \\ 
       E-mail: dmameyer.astro@gmail.com ; meyer@ice.csic.es  \\ 
       }

\begin{document}

\date{Received; accepted}

\maketitle

\label{firstpage}

\begin{abstract} 
Cosmic rays are highly energetic messengers propagating in magnetized plasma,  
which are, possibly but not exclusively, accelerated at astrophysical shocks. 
Amongst the variety of astrophysical objects presenting shocks, the huge circumstellar stellar wind bubbles forming around very massive stars, are potential non-thermal emitters. We present the 1D magneto-hydrodynamical simulation of the evolving magnetized surroundings of a single, OB-type 
main-sequence $60\, \rm M_\odot$ star, which is post-processed to calculate the re-acceleration 
of pre-existing non-thermal particles of the Galactic cosmic ray background. It is found that the 
forward shock of such circumstellar bubble can, during the early phase ($1\, \rm Myr$) of its 
expansion, act as a substantial re-accelerator of pre-existing interstellar cosmic rays. This 
results in an increasing excess emission ﬂux by a factor of 5, the hadronic component producing $\gamma$-rays by $\pi^0$ decay being more important than those by synchrotron and inverse 
Compton radiation mechanisms. We propose that this eﬀect is at work in the circumstellar environments 
of massive stars in general and we conjecture that other nebulae such as the stellar 
wind bow shocks of runaway massive stars also act as Galactic cosmic-ray re-accelerators. 
%
%
Particularly, this study supports the interpretation of the enhanced hadronic emission ﬂux measured from the surroundings of $\kappa$ Ori as originating from the acceleration of pre-existing particles at the forward shock of its wind bubble.
\end{abstract}



\section{Introduction}
\label{sect:intro}

The detailed components of the non-thermal spectrum of the interstellar 
medium (ISM) of the Milky Way is far from being understood. 
A source of Galactic cosmic rays is the so-called remnants nebulae, 
produced by the fast expansion of a supernova blastwave, released 
by the death of a star or a stellar system, into its local ambient 
medium~\citep{shklovskii_1954,weiler_araa_25_1988,blasi_crpa_conf_2011, 
vink_aarv_20_2012,vink_2020}. 
Such cosmic rays, principally emitting in radio, X-rays, and $\gamma$-rays 
via inverse Compton, non-thermal synchrotron mechanisms 
and hadronic processes, are well 
documented thanks to the plethora of data collected by high-energy 
observatories such as \textit{Fermi} and \textit{H.E.S.S.}, which permitted 
to confront theoretical models like the diffusive shock 
acceleration~\citep{schatzman_anap_26_1963,axford_icrvc_1977, 
bell_mnras_182_1978a,bell_mnras_182_1978b, blandford_apj_221_1978, 
krymsky_ICRC_1979} with observations~\citep{abdo_apj_722_2010,hess_2018AA_612A_8H}. 
The forthcoming \textit{Cherenkov Telescope Array (CTA} ground-based observatory 
will continue this ongoing effort \citep{acharyya_mnras_523_2023,acero_cta_2023}. 
However, if shocks and turbulence in supernova remnants are 
identified accelerators of charged particles, at least in the sub-PeV 
band of the cosmic ray spectrum, additional unidentified sources 
able to bring electrons and protons of the ISM to very high kinetic 
energies must exist to fully explain the Galactic cosmic ray 
spectrum before and beyond the knee~\citep{cristofari_aa_650_2021}. 
Amongst all potential cosmic-ray accelerators, wind collision in binary 
systems~\citep{reimer_apss_309_2007,hamaguchi_natas_2_2018,pittard_mnras_504_2021}, 
pulsars nebulae~\citep{bednarek_aa_423_2004} but also superbubbles 
around stellar clusters 
with massive stars~\citep{bykov_ssrv_99_2001,butt_apj_677_2008,  
morlino_mnras_504_2021,2021arXiv210801870M} 
have been theoretically demonstrated to 
present the ability to substantially accelerate particles  
within the diffusive shock acceleration, which turned out to be 
consistent with high-energy 
observations~\citep{joubaud_aa_635_2020}. 
Particularly, there is a growing suspicion for the circumstellar shocks 
found in nebulae generated by the interaction between the strong winds of 
high-mass stars and the ISM, to be a major source of Galactic cosmic 
rays~\citep{cardillo_galax_7_2019,meyer_mnras_493_2020}.

The basic picture of the circumstellar medium of massive stars is the 
wind-ISM interaction model of~\citet{weaver_apj_218_1977}, in which the 
supersonic wind of a main-sequence massive star collides with its constant 
local ambient medium. 
This results in a structured nebula, delimited by an inner termination 
shock (the so-called reverse shock) and an outer so-called forward shock, 
and centered onto the region of free-steaming stellar wind. 
The gas between the reverse and forward shocks is made of two layers of 
shocked diluted hot stellar wind and shocked dense cold ISM material, 
separated by a contact discontinuity.  
Those simple objects motivated numerous studies, from the beginning of 
numerical fluid dynamics to recent computing-intensive calculations, 
revealing their complex functioning, moderated by radiative 
cooling~\citep{vanmarle_comp&fluids_2010pdf}, thermal 
conduction~\citep{zhekov_na_3_1998}, stellar 
evolution~\citep{avedisova_saj_15_1972,garciasegura_aa_455_1995a, 
garciasegura_aa_455_1995b,garciasegura_1996_aa_305f} 
and their sensitivity to the large-scale magnetization of the 
ISM~\citep{vanmarle_584_aa_2015}. 
If the star moves fast through the ISM, the wind bubble is turned into 
a bow shock~\citep{gull_apj_230_1979,kobulnicky_apj_710_2010,2012A&A...537A..35C, 
Gvaramadze_2013,kobulnicky_aj_154_2017}, characterized by an 
arc-like shape, an internal organization similar to circumstellar wind bubbles, 
and by a radiative forward shock that further stimulated the analytic and 
numerical exploration of the surroundings of massive runaway 
stars~\citep{wilkin_459_apj_1996,comeron_aa_338_1998,vanmarle_apj_734_2011, 
meyer_obs_2016,acreman_mnras_456_2016}. 
Stellar wind bubbles and bow shocks have been largely studied as the host site  
where the most massive stars die as a core-collapse supernova 
event~\citep{dwarkadas_apj_630_2005,dwarkadas_apj_667_2007}, 
giving birth to complex supernova remnants~\citep{vanveelen_aa_50_2009, 
meyer_mnras_450_2015,meyer_mnras_502_2021,velazquez_mnras_519_2023,
villagran_mnras_527_2024} or $\gamma$-ray 
bursts~\citep{vanmarle_aa_444_2005,vanmarle_aa_460_2006,vanmarle_aa_469_2007}.  
Nevertheless, their role as cosmic-rays accelerators still deserve  
investigations.

The Earth's closest circumstellar structure is the termination shock of the 
solar wind interacting with the ISM. It has been theoretically
suggested~\citep{jokipii_apj_152_1968,jokipii_apj_611_2004} and interpreted in
the context of measures of the spacecraft {\sc Voyager} to produce a local 
component of cosmic rays~\citep{ferreira_adspr_41_2008,jopikii_apj_794_2014, 
pogorelov_ssrv_212_2017}. 
This idea has been extended to the termination (reverse) shock of circumstellar 
wind bubbles around main-sequence, OB-type massive stars~\citep{casse_apj_237_1980}, 
although this results have been questioned~\citep{voelk_apj_253_1982} but also  
confirmed~\citep{webb_apj_298_1985}. 
Equivalently, the raising number of detected stellar wind bow shocks of 
supersonically-moving high-mass stars~\citep{peri_aa_538_2012,peri_aa_578_2015,   
kobulnicky_apjs_227_2016,kobulnicky_apj_856_2018} spurred the interest of 
the heliospheric community to massive stellar objects such as the OB star $\lambda$ 
Cephei, predicting local cosmic-ray anisotropies to be generated inside of  
such stellar wind cavities~\citep{scherer_aa_576_2015,herbst_ssrv_218_2022}.

The observational and experimental quest for cosmic rays originating 
from the surroundings of massive stars is a vivid field of research. 
On the one hand, the astrospheres of runaway massive stars are predicted 
to be the site of particle acceleration~\citep{delvalle_aa_543_2012, 
delvalle_aa_563_2014,valle_mnras_448_2015,valle_ApJ_864_2018}. 
Such emissions have been detected from the bow shock EB27 around 
BD+43${\degree}$3654~\citep{benaglia_aa_517_2010, 
palacio_aa_617_2018,benaglia_mnras_503_2021} and 
possible association 
with $Fermi$ $\gamma$-rays data have been reported in~\citet{sanchezayaso_apj_861_2018}.  
Recently, a second stellar wind bow shock from a massive runaway 
star revealed in its turn radio non-thermal signature~\citep{2022arXiv220411913M}.
On the other hand, more systematic attempts to measure, e.g. high-energy 
X-rays~\citep{schultz_aa_565_2014,toala_apj_821_2016,tola_apj_838_2017,
DeBecker_mnras_471_2017,binder_aj_157_2019} and very 
high-energy emission~\citep{schultz_aa_565_2014} from the surroundings of 
stellar wind bow shocks turned to be unfruitful. 
At the same time, the surroundings of evolved Wolf-Rayet 
stars seem to be non-thermal emitters~\citep{prajapati_apj_884_2019}, 
as predicted by~\citet{zirakashvili_aph_98_2018} and phenomenological studies 
pointing out the importance of cosmic-rays re-acceleration in the 
circumstellar shocks~\citep{cardillo_galax_7_2019}. 
This puzzling situation leaves wide open the question of the existence of 
cosmic rays from circumstellar shock.

This paper is a numerical investigation of the non-thermal phenomenon 
in the context of the circumstellar medium of a very massive 
($\ge 60\, \rm M_\odot$) star. 
%
%
%
%
We explore, by means of 1D magneto-hydrodynamical (MHD) simulations completed 
by particle acceleration calculations, the possibility of expanding stellar 
wind bubbles around a very massive star to generate non-thermal emission
by the re-acceleration of pre-existing Galactic background cosmic rays. 
We established time-dependent predictions for the non-thermal synchrotron 
radiation, inverse Compton emission, and hadronic $\gamma$-rays feedback 
by $\pi^{o}$ decay of the surroundings of a very massive star, which we 
further consider in the context of observational data.

Our study is organized as follows. We first introduce the reader to the various 
numerical methods that we have used to simulate the 1D MHD stellar wind bubble 
of a $60\, \rm M_\odot$ star and to calculate its non-thermal emission in 
Section~\ref{sect:method}. 
In Section~\ref{sect:results} we present results for the MHD structure of the 
surroundings of very massive stars, the particle acceleration at the shocks 
generated by wind-ISM interaction, and the resulting non-thermal emission. 
The caveats of our results are further detailed in 
Section~\ref{sect:discussion}, discussed in the context of observations and  
we conclude in Section~\ref{sect:conclusion}.


\section{Methods}
\label{sect:method}

In this section, we present the numerical methods used to model the magnetized 
circumstellar medium of a very massive star and to calculate the acceleration 
of particles at work therein.

\subsection{MHD stellar wind bubbles}
\label{sect:pluto}

\subsubsection{Governing equations}
\label{sect:gov_eq}

The problem of the plasma in MHD circumstellar bubbles of massive stars is described 
as follows by the non-ideal MHD equations, 
\begin{equation}
	   \frac{\partial \rho}{\partial t}  + 
	   \bmath{\nabla}  \cdot \big(\rho\bmath{v}\Big) =   0,
\label{eq:mhdeq_1}
\end{equation}
\begin{equation}
	   \frac{\partial \bmath{m} }{\partial t}  + 
           \bmath{\nabla} \cdot \Big( \bmath{m} \otimes \bmath{v} 
           - \bmath{B} \otimes \bmath{B} + \bmath{\hat I}p \Big) 
            =   \bmath{0},
\label{eq:mhdeq_2}
\end{equation}
\begin{equation}
	  \frac{\partial E }{\partial t}   + 
	  \bmath{\nabla} \cdot \Big( (E+p)\bmath{v}-\bmath{B}(\bmath{v}\cdot\bmath{B}) \Big)  
	  = \Phi(T,\rho),
\label{eq:mhdeq_3}
\end{equation}
and,
\begin{equation}
	  \frac{\partial \bmath{B} }{\partial t}   + 
	  \bmath{\nabla} \cdot \Big( \bmath{v}  \otimes \bmath{B} - \bmath{B} \otimes \bmath{v} \Big)  =
	  \bmath{0},
\label{eq:mhdeq_4}
\end{equation}
with $\rho$ the gas density, $\bmath{v}$ the plasma velocity vector, 
$p$ the thermal pressure of the gas, $\bmath{\hat I}$ the identity matrix, 
\begin{equation}
    \bmath{m}=\rho\bmath{v},
\end{equation}
the linear momentum vector, $\bmath{B}$ the magnetic field vector and 
\begin{equation}
	E = \frac{p}{(\gamma - 1)} + \frac{ \bmath{m} \cdot \bmath{m} }{2\rho} 
	    + \frac{ \bmath{B} \cdot \bmath{B} }{2},
\label{eq:energy}
\end{equation}
the total energy of the system. This system of equations is closed by the 
definition of the adiabatic sound speed in the medium, 
\begin{equation}
	c_{\rm s} = \sqrt{ \frac{\gamma p }{\rho} },
\label{eq:cs}
\end{equation}
with $\gamma=5/3$ the adiabatic index for ideal gas.

Losses by optically-thin radiative processes are represented by the term,  
\begin{equation}  
	 \Phi(T,\rho)  =  n_{\mathrm{H}}\Gamma(T)   
		   		 -  n^{2}_{\mathrm{H}}\Lambda(T),
\label{eq:dissipation}
\end{equation}
with, 
\begin{equation}
	T =  \mu \frac{ m_{\mathrm{H}} }{ k_{\rm{B}} } \frac{p}{\rho},
\label{eq:temperature}
\end{equation}
the gas temperature where $k_{\rm{B}}$ is the Boltzmann constant, $m_{\rm H}$ the 
mass of the proton, $\mu$ is mean molecular weight, set to $0.61$ for gas ionized 
by the radiation of hot massive stars. 
The function $\Lambda (T)$ stands for the cooling rate, while $\Gamma(T)$ 
represents the heating rate of the gas, respectively. 
We refer \textcolor{black}{the reader} interested to the detailed description of the $\Gamma(T)$ and 
$\Lambda(T)$ to~\citet{meyer_2014bb}.

The MHD simulations of circumstellar bubbles of massive stars are carried 
out with the {\sc pluto} code~\citep{mignone_apj_170_2007, migmone_apjs_198_2012,vaidya_apj_865_2018} 
using the Harten-Lax-van Leer Riemann solver~\citep{hll_ref} with a 
$2^{\rm nd}$-order numerical scheme together with linear reconstruction 
of the primitive variables. 
We make use of the eight-wave MHD formulation by~\citet{Powell1997} which ensures  
\textcolor{black}{that the} divergence-free condition of the magnetic field vector, 
\begin{equation}
    \bmath{\nabla} \cdot \bmath{B} = \bmath{0}, 
\label{eq:div_B}
\end{equation}
is satisfied everywhere in the computational domain, throughout the 
whole calculation.  
Last, the timestep of the numerical simulations is controlled by the 
Courant-Friedrich-Levy criterion, which we set initially to $C_{\rm cfl}=0.1$.

\subsubsection{Initial conditions}
\label{sect:ic}

We perform our MHD wind bubble models using a 1D spherically-symmetric 
coordinate system and a $[0,r_{\rm rmax}]$ computational domain with 
$r_{\rm max}=150\, \rm pc$, that is mapped with a uniform grid of 
$2000$ zones. To avoid spurious MHD effects at the stellar wind 
boundary, we first model the early $0.01\, \rm Myr$ of the wind-ISM 
interaction in a pure hydrodynamical manner and using a smaller 
domain of $r_{\rm max}=15\, \rm pc$ uniformly discretised with $1000$ 
zones, before mapping the solution onto the larger domain. 
As in~\citet{dwarkadas_apj_630_2005,dwarkadas_apj_667_2007}, we perform 
the simulations in the reference frame of the star. 
The stellar wind is launched within a spherical zone of $20$ grid zones 
centered onto the origin of the computational domain. The wind density 
is as follows, 
\textcolor{black}{
\begin{equation}
	\rho_{w}(r,t) = \frac{ \dot{M}(t) }{ 4\pi r^{2} v_{\rm w}(t) },
    \label{eq:wind}
\end{equation}
with $r$ the distance to the star, $\dot{M}(t)$ the mass-loss rate and 
$v_{\rm w}(t)$ the time-dependent stellar wind terminal velocity. 
The various surface properties of the star ($\dot{M}(t),v_{\rm w}(t)$)} are 
taken from the tabulated stellar evolutionary track for the $60\, \rm M_\odot$ star 
of~\citet{groh_aa564_2014}, which has already been used to simulate wind 
bubbles a two-dimensional fashion~\citep{meyer_mnras_493_2020}. 
The ambient medium has properties corresponding to the warm phase of the 
ISM, with constant number density $n_{\rm ISM}=0.79\, \rm cm^{-3}$ 
and temperature $8000\, \rm K$, respectively.

Particle acceleration being of intrinsic MHD nature, a magnetic field is added 
into the outflowing stellar wind. 
The magnetic field is incorporated into the model at the time $0.01\, \rm Myr$, 
at the moment of the mapping of the hydrodynamical wind-ISM interaction to 
the MHD grid. 
The stellar boundary conditions are made of a radial component, 
\begin{equation}
	B_{\rm r}(r) = B_{\star} \Big( \frac{R_{\star}}{r} \Big)^{2},
    \label{eq:Br}
\end{equation}
and of a toroidal component, 
\begin{equation}
	B_{\phi}(r) =  B_{\rm r}(r) 
	\Big( \frac{ v_{\phi} }{ v_{\rm w} } \Big) 
	\Big( \frac{ r }{ R_{\star} }-1 \Big),
    \label{eq:Bphi}
\end{equation}
where $B_{\star}$ is the stellar surface magnetic field, $R_{\star}$ 
the stellar radius and $v_{\phi}=v_{\rm rot}$ is the toroidal stellar 
rotation velocity at the equator~\citep{meyer_mnras_506_2021,meyer_mnras_507_2021}. 
Since we perform 1D MHD simulations, any latitude-dependence of the stellar 
wind has disappeared and we calculate the models in the \textcolor{black}{equatorial plane} of 
the spherical coordinate system ($r$,$\theta=\pi/2$,$\phi$). 
In this study, we adopt the stellar surface properties of 
$B_{\star}=50\, \rm G$, and $v_{\rm rot}=50\, \rm km\, \rm^{-1}$, 
which \textcolor{black}{are characteristic} of OB stars~\citep{hubrig_aa_551_2013}. 
Values for the stellar radius $R_{\star}$ are interpolated from the 
track of~\citet{groh_aa564_2014}. 
Magnetized stellar wind \textcolor{black}{boundary conditions} of MHD circumstellar
simulations are further described in~\citet{chevalier_apj_421_1994, 
rozyczka_apj_469_1996,garciasegura_apj_860_2018,herbst_apj_897_2020, 
garciasegura_apj_893_2020,baalmann_aa_650_2021,meyer_mnras_506_2021, 
2021arXiv210809273M}.

Inside of the computational domain, we impose a 
reconstructed magnetic field, taking the form of a piece-wise 
function. 
In the remaining part of the paper, we will note the termination 
shock $R_{\rm R}$, the contact discontinuity $R_{\rm CD}$ and the 
forward shock $R_{\rm F}$.
The freely-expanding wind region ($r \le R_{\rm R}$) 
and the shocked stellar wind ($R_{\rm R} \le r \le R_{\rm CD}$), 
are filled with, 
\begin{align}
 B_{\rm r}  & = B_{\star} \Big( \frac{R_{\star}}{r} \Big)^{2}      \nonumber \\
 B_{\theta} & = 0      \nonumber \\ 
 B_{\phi}   & = B_{\rm r}(r) \Big( \frac{ v_{\phi} }{ v_{\rm w} } \Big) 
             	\Big( \frac{ r }{ R_{\star} }-1 \Big), 
\end{align}
while the region of shocked ISM ($R_{\rm CD} \le r \le R_{\rm F}$) is 
filled as, $B_{\rm r}=B_{\theta}=B_{\phi}=\sigma B_{\rm ISM}/\sqrt{3}$, 
where $\sigma$ is the compression ratio of the shock, taken to $\sigma=4$. 
Hence, the value of the compressed ISM magnetic field in the outer layer 
of the wind bubble is, 
\begin{align}
	B(R_{\rm CD}\le r \le R_{\rm F})  
	& = \sqrt{ \Big( B_{\rm r} \Big)^{2} 
	    + \Big( \sigma B_{\theta} \Big)^{2} 
	    + \Big( \sigma B_{\phi} \Big)^{2} } 
	  \nonumber \\
	& = \sqrt{ 1 + 2\sigma^{2} } B_{\rm ISM},  	 
    \nonumber \\
	& = \sqrt{ 11 } B_{\rm ISM},  
\end{align}
with $B_{\rm r}=B_{\theta}=B_{\phi}=B_{\rm ISM}/\sqrt{3}$ 
and $B_{\rm ISM}=7 \mu\, \rm G$.
Similarly, we consider the magnetization of the ISM,  
\begin{equation}
	B(r>R_{\rm F}) 
	= \sqrt{ B_{\rm r}^{2} + B_{\theta}^{2} + B_{\phi}^{2} }=B_{\rm ISM},
\end{equation}
also with $B_{\rm r}=B_{\theta}=B_{\phi}=B_{\rm ISM}/\sqrt{3}$
in each grid zone of the computational domain $R_{\rm F}<r$. 
After the mapping, the 1D MHD simulations are restarted at a time $t=0.01$ 
and further calculated over the part of the main-sequence phase for which 
the compression ratio of the forward shock propagating into the ISM is 
larger than 4. This corresponds to the early $1\, \rm Myr$ of the star's 
evolution~\citep{groh_aa564_2014} and we temporally discretize our 
simulation so that about $100$ MHD simulation outputs are produced.

\begin{figure*}
        \centering
        \includegraphics[width=1.0\textwidth]{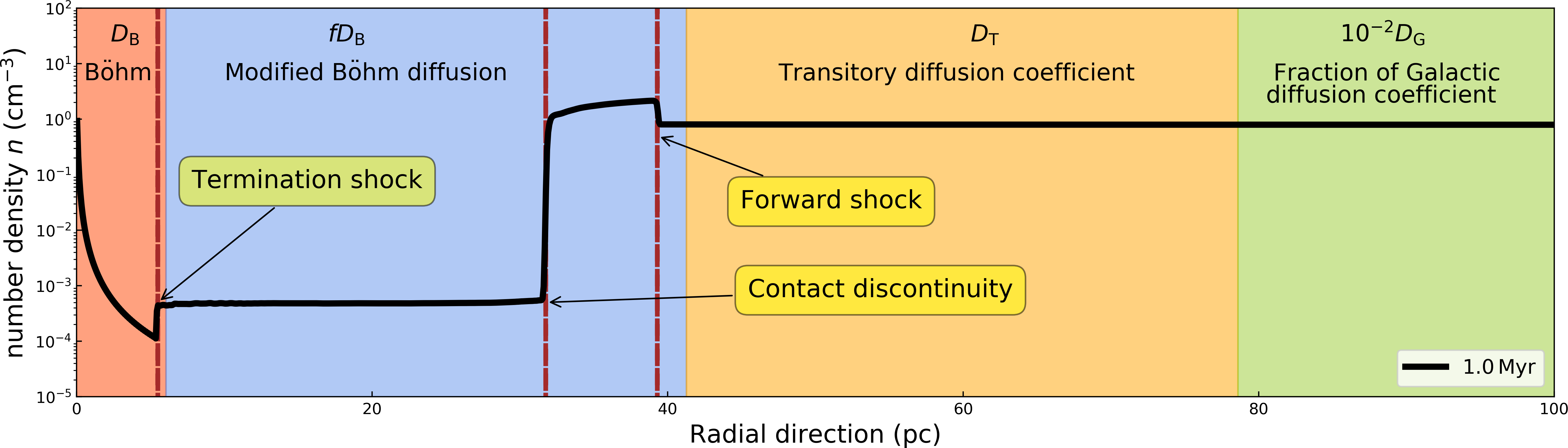}
        \caption{
        Number density of a $1\, \rm Myr$ old stellar wind bubble, 
        formed by a $60\, \rm M_\odot$ 
        star (black lines in $\rm cm^{-3}$). 
        The figure highlights 
        its principal discontinuities (termination shock, contact 
        discontinuity, forward shock, dashed red lines), but also 
        how the B\"ohm-like diffusion coefficient $D(r)$ of 
        Eq.~\ref{eq:coeff_diff} is treated as a function of 
        the distance to the star, into the different regions 
        of the circumstellar medium. 
        }
        \label{fig:sketch}  
\end{figure*}

\subsection{Particle acceleration calculations}
\label{sect:ratpac}

The 1D MHD stellar wind bubble model simulated with the {\sc pluto} code 
\textcolor{black}{is} post-processed with the particle acceleration code {\sc ratpac}. 
The Radiation Acceleration Transport Parallel Code ({\sc ratpac}) is a 
numerical framework solving the acceleration and subsequent transport of 
electrons and protons in a magnetized plasma presenting shocks. 
It is a modular, parallelized code written in {\sc python} and \textcolor{black}{making 
use} of the {\sc fipy} library of finite volume solvers for partial 
differential equations. 
Originally developed in the context of supernova remnants, {\sc ratpac} 
\textcolor{black}{post-processes} arbitrary pre-calculated or analytic plasma flows including 
a pattern of MHD shocks~\citep{telezhinsky_aph_35_2012, 
telezhinsky_aa_541_2012,telezhinsky_aa_552_2013,wilhelm_aa_639_2020}, 
although it can self-consistently be used on the fly, coupled together 
with an MHD solver such as the {\sc pluto} 
code~\citep{brose_aa_593_2016,sushch_aa_618_2018, 
2021arXiv210810773B}.

%

The MHD profiles ($\rho$, $T$, $\bmath{v}$ $\bmath{B}$) are loaded into 
{\sc ratpac} and, at each timestep of the calculation, the MHD flow 
\textcolor{black}{profile is} interpolated between the temporal neighboring snapshots. 
The coordinate transformation, 
\begin{equation}
	( x - 1 ) = ( x^{*} - 1 )^{3},
	\label{coord_transf}
\end{equation}
with $x=r/R_{\rm F}$, permits us to greatly reduce numerical 
costs and to reach spatial resolution near the shock region much finer 
than that of the pre-calculated MHD profiles. 
Once the interpolated MHD profile is transformed to the new coordinate system, 
a shock finder algorithm looks for the position of the forward shock of the 
stellar wind bubble $R_{\rm F}$. 
The position of the shock is found by looping through the radial direction, 
from the ISM to the star, $R_{\rm F}$ being caught via the compression 
ratio $\sigma \ge 2$ of the density. To compensate for the poor resolution 
in the vicinity of the shock, the velocity profile is fitted with a logistic
function using 5 grid zones upstream and downstream of the forward shock. 
The velocity flow is finally resharpened to ensure a clear sharp jump around 
$R_{\rm F}$. 
%

The time-dependent numerical method is based on kinetic calculations
within the test-particle approximation, which neglects any feedback of 
the cosmic rays to the gas dynamics. 
The main output of {\sc ratpac} is the non-thermal particle number 
density $N(r,p,t)$ with $r$ the radial coordinate, $p$ the momentum 
(or equivalently the energy $E$) of the particle and $t$ the time, respectively. 
The evolution of $N$ is determined in spherical coordinates, 
with a grid that is co-moving with the considered shock, where 
the following diffusion-advection equation is solved, 
\begin{equation}
	\frac{ \partial N }{ \partial t } = 
	\bmath{\nabla} \Big( D \bmath{\nabla} N - \bmath{v} N \Big) 
	- \frac{ \partial} { \partial t } 
	\Big[ N \frac{ \partial p }{ \partial t } - \frac{ \bmath{\nabla} N \cdot 
	\bmath{v} }{ 3 }  \Big] , 
\end{equation}
where $D$ is the spatial diffusion coefficient, $\bmath{v}$ is the vector 
velocity, $\partial p/\partial t$ represents the energy losses.

\subsection{Diffusion model}
\label{sect:diffusion}

\textcolor{black}{
A B\" ohm-like diffusion model is adopted 
in this study, controlled by a free parameter $f$. 
}
We adopt a coefficient that is a piece-wise function of the distance 
to the central star~\citep{telezhinsky_aph_35_2012, 
telezhinsky_aa_541_2012,telezhinsky_aa_552_2013,wilhelm_aa_639_2020}, 
which includes a free parameter $f$ accounting 
for sub-grid microphysical diffusion processes at work in the interior of the 
circumstellar bubble, such as magnetic turbulence cascade, not included 
in our simulations. 
%
It reads, 
\begin{align}
D(r) = \left\{ \begin{array}{lcc} 
                D_{\rm B}         & \hspace{5mm} 0 \le r \le 1.1 R_{\rm F}  \\
                fD_{\rm B}        & \hspace{5mm} 1.1 R_{\rm F} \le r \le 1.05 R_{\rm F} \\
                D_{\rm T}(r)      & \hspace{5mm} 1.05 R_{\rm F} \le r \le 2.0 R_{\rm F} \\
                10^{-2}D_{\rm G}  & \hspace{5mm} 2.0 R_{\rm F} \le r \\                
                \end{array} \right.
\label{eq:coeff_diff}                
\end{align}
where 
\begin{equation}
	D_{\rm G} = D_{0} \Big( \frac{E}{10\, \rm GeV}  \Big)^{\alpha} 
	                  \Big( \frac{B}{3\, \mu \rm G} \Big)^{-\alpha}, 
\end{equation}
is the diffusion coefficient of the Galactic cosmic ray background 
acting as a bath of non-thermal particles 
in which the wind bubble evolves, with $\alpha=1/3$ and $D_{0}=10^{29}\, 
\rm cm^{2}\, \rm s^{-1}$, and where,  
\begin{equation}
	D_{\rm T}(r) = D_{\rm B} \exp 
	              \Big[ 
	              \ln \Big( \frac{0.01 D_{\rm G}}{D_{\rm B}}  \Big) 
	              \frac{  (r-1.05 R_{\rm F})  }{  (2-1.05)R_{\rm F}  }
	              \Big], 
\end{equation}
represents the exponential transition between the B\" ohm-based diffusion coefficient 
in the bubble and that in the ISM. Our work explores the effects of the diffusion 
coefficient by varying it in the range $10\le f \le 400$. 
Our diffusion coefficient strategy is further illustrated in Fig.~\ref{fig:sketch}.

\subsection{Energy threshold and particle re-acceleration}
\label{sect:reacc}

We include the re-acceleration of Galactic cosmic rays into the simulations by 
switching-off the injection of particles~\citep{wilhelm_aa_639_2020} and by imposing 
a reservoir of non-thermal particle distribution taken the Galactic cosmic-ray 
background spectrum. The non-thermal particles are initially distributed 
using a uniform spatial distribution from the outer boundary of the 
computational domain up to the termination shock of the stellar wind bubble. 
The initial time of the particle acceleration 
simulation is performed when the stellar wind bubble is $0.5\, \rm Myr$ old, 
time instance from which the termination shock, contact discontinuity, and forward 
shock of the circumstellar bubble, is clearly distinguishable and capturable 
by the shock/discontinuity-finder algorithm. 
Neglecting the early phase of the bubble's expansion underestimate ISM particle 
re-acceleration, potentially important as the compression ratio of the forward 
shock is large. The goal of this work is not to estimate the total, precise non-thermal 
budget of the wind bubble, but rather to qualitatively corroborate previous observations, 
see Section \ref{sect:discussion}.

The cosmic-ray spectrum that we adopt is that of~\citet{brose_aa_593_2016} 
and~\citet{wilhelm_aa_639_2020}, modified for the purpose of the present 
study: since the Galactic cosmic rays do not reach the termination shock 
of wind bubbles around massive stars~\citep{voelk_apj_253_1982}, the 
pre-existing non-thermal particles are imposed from the ISM to the 
contact discontinuity of the bubble $R_{\rm CD}$. It reads, 
\begin{align}
N_{\rm Gal}(r) = 
\left\{
    \begin{array}{lr}
        0,               & r  \le R_{\rm CD} \\
        N_{\rm Gal}(E),  & r  >   R_{\rm CD}
    \end{array}\right.
\end{align}
where $N_{\rm Gal}(E)$ is taken from~\citet{wilhelm_aa_639_2020} for the 
electrons, and from~\citet{Moskalenko_apj_565_2002} 
and~\citet{Jaffe_mnras_416_2011} for the protons, respectively.  
To be re-accelerated at the shock by diffusive 
shock acceleration, the particles need to have energy sufficiently high to 
cross the shock the first time and begin the Fermi cycles. We call $E_{\rm th}$ the 
threshold energy necessary to trigger the re-acceleration process of a particle.  
This lower-energy cut-off of the distribution of available cosmic ray is  
the governing parameter of diffusive 
re-acceleration~\citep{thornbury_mnras_442_2014,drury_icrc_34_2015}. 
We introduce a corresponding cut-off in the Galactic cosmic-ray background 
in order not to artificially take into account the numerous low-energy particles 
which would bias the solution by being considered as accelerable by the code 
while they can not cross the forward shock.

\textcolor{black}{
The quantity $E_{\rm th}$ is evaluated by forcing the pre-existing non-thermal 
particles at the forward shock to have properties such that their  
acceleration characteristic lengthscale (the precursor length $L_{\rm acc}$) 
is larger than the shock's radiative characteristic lengthscale 
(the cooling length $L_{\rm cool}$). 
}
They read, 
\begin{equation}
	L_{\rm cool} = v t_{\rm cool} = \frac{ v_{\rm F} }{ \kappa } t_{\rm cool},
	\label{lcool}
\end{equation}
and 
\begin{equation}
	L_{\rm acc} = \frac{ f D_{\rm B} }{ v_{\rm F} },
	\label{lacc}
\end{equation}
respectively, with $v_{\rm F}$ the speed of the forward shock expanding through 
the ISM and $v$ the gas velocity in the post-shock region at the forward shock. 
In the above relation, $t_{\rm cool}$ is the cooling timescale,  
\begin{equation}
	t_{\rm cool} = \frac{ k_{\rm B} T_{\rm d} n_{\rm d} }{ 
	               (\gamma-1) \Lambda(T_{\rm d}) n_{\rm d}^{2} }
	             = \frac{ k_{\rm B} T_{\rm d}  }{ 
	               (\gamma-1) \Lambda(T_{\rm d}) \kappa n_{\rm u} }    ,
	\label{tcool}
\end{equation}
as defined in~\citet{meyer_mnras_464_2017}, with $n_{\rm u}=n_{\rm ISM}$ 
the number density upstream the shock, $n_{\rm d}=\kappa n_{\rm u}$ 
the number density downstream of the magnetized forward shock of the 
bubble measured from the simulation, and with, 
\begin{equation}
	\kappa = \frac{  2(\gamma+1)  }{ 
	\Delta + \sqrt{   \Delta^{2} + \frac{4(\gamma+1)(2-\gamma)}{M_{\rm A}}  }  },
	\label{kappa}
\end{equation}
the compression ratio at the forward shock, and,
\begin{equation}
	\Delta = (\gamma-1) \Big(   \frac{2}{M^{2}} + \frac{\gamma}{M_{\rm A}^{2}}  \Big),
	\label{kappa}
\end{equation}
where $M$ and $M_{\rm A}$ \textcolor{black}{are the Mach and Alfvenic Mach} numbers upstream the 
shock, respectively~\citep{shu_pavi_book_1992}. 
Similarly, the temperature downstream of the shock,  
\begin{equation}
	T_{\rm d} = T_{\rm ISM} \frac{ ((\gamma-1)M^{2}+2)(2\gamma M^{2}-(\gamma-1)) }
	            { (\gamma+1)^{2} M^{2} },
	\label{Td}
\end{equation}
is the gas temperature downstream the shock, while $T_{\rm ISM}=8000\, \rm K$ is the 
upstream shock temperature, and $\Lambda(T)$ is the rate for 
optically-thin cooling for ionized gas that is interpolated from the cooling
curve presented in~\citet{meyer_2014bb}. The B\" ohm diffusion coefficient 
reads, 
\begin{equation}
	D_{\rm B} = \frac{ pc }{ 3eB_{\rm d} },
	\label{Td}
\end{equation}
with $c$ the speed of light, $e$ is the elementary charge, 
and $B_{\rm d}=\sqrt{1+2\kappa^{2}/3}=B_{\rm ISM}/\sqrt{3}$ 
the compressed magnetic field downstream of the forward shock.

\begin{figure}
        \centering
        \includegraphics[width=0.48\textwidth]{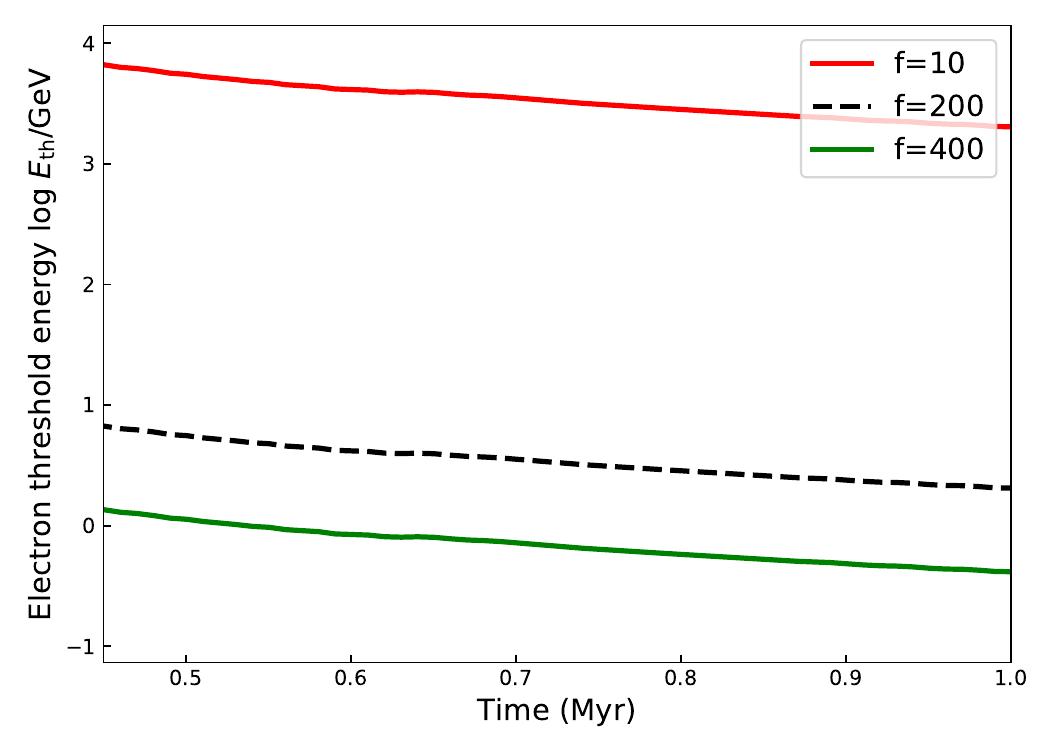}        
        \includegraphics[width=0.48\textwidth]{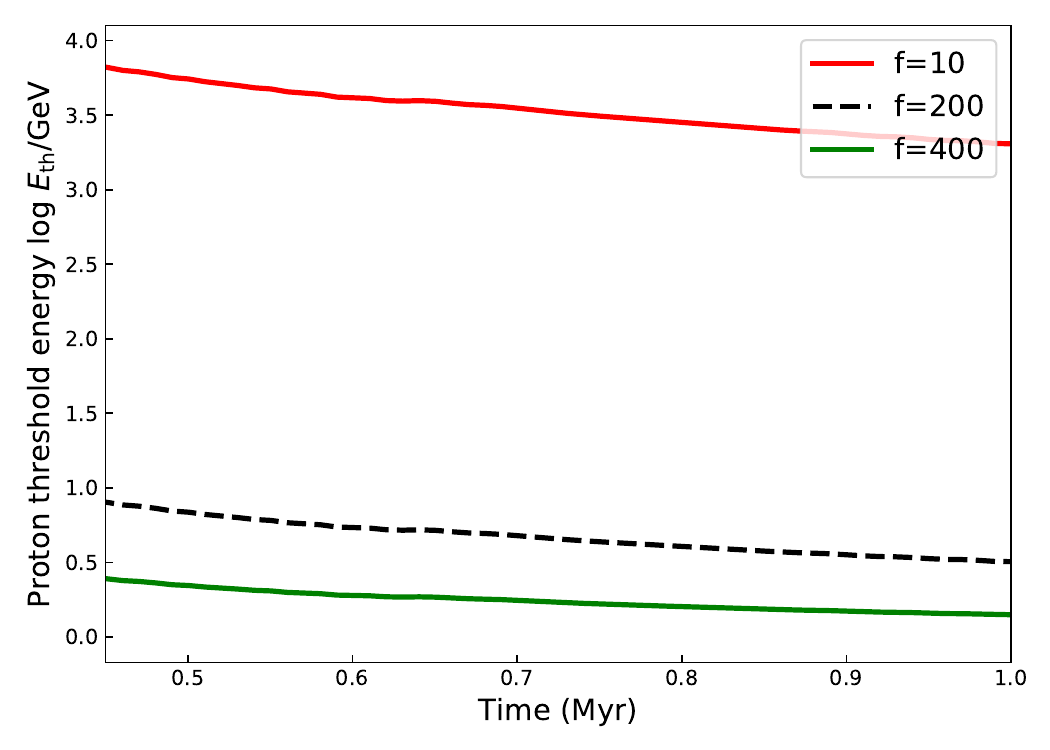}
        \caption{
        Time evolution of the energy threshold $E_{\rm th}$ used to cut 
        off the Galactic cosmic-ray background, for electrons (a) 
        and for protons (b). 
        The quantity are plotted as a function of time (in $\rm Myr$) 
        for different multiplying factor $f$ of the B\" ohm diffusion 
        coefficient $D_{\rm B}$. 
        }
        \label{fig:plot_Eth}  
\end{figure}

The value of the cut-off momentum $p_{\rm th}$ is obtained by solving 
$L_{\rm cool}=L_{\rm acc}$ for $p$, which straightforwardly provides 
the cut-off energy $E_{\rm th}$, for both electrons and protons. 
The evolution of $E_{\rm th}$ for several values of $f$ is plotted in 
Fig.~\ref{fig:plot_Eth}. It shows that the cut-off energy is of the same 
order of magnitude during an early $1\, \rm Myr$ of the star's life, 
and decreases with time as the wind bubble expands and the compression ratio 
of its forward shock gets smaller. 
%
%
%
Our results are consequently a \textcolor{black}{lower limit estimate} of the non-thermal feedback 
for the outer circumstellar shocks around early-type, main-sequence massive stars. 
Without including this initial non-thermal population in the ISM, the process of 
accelerating particles at shocks during the wind bubble initial expansion 
phase is inefficient and its associated emission is consequently negligible. 
This capability to accelerate particles decreases as the bubble further 
grows
and its compression ratio decreases, as accelerated protons 
escape and as the electrons lose energy by radiation 
mechanisms.

\begin{figure}
        \centering
        \includegraphics[width=0.48\textwidth]{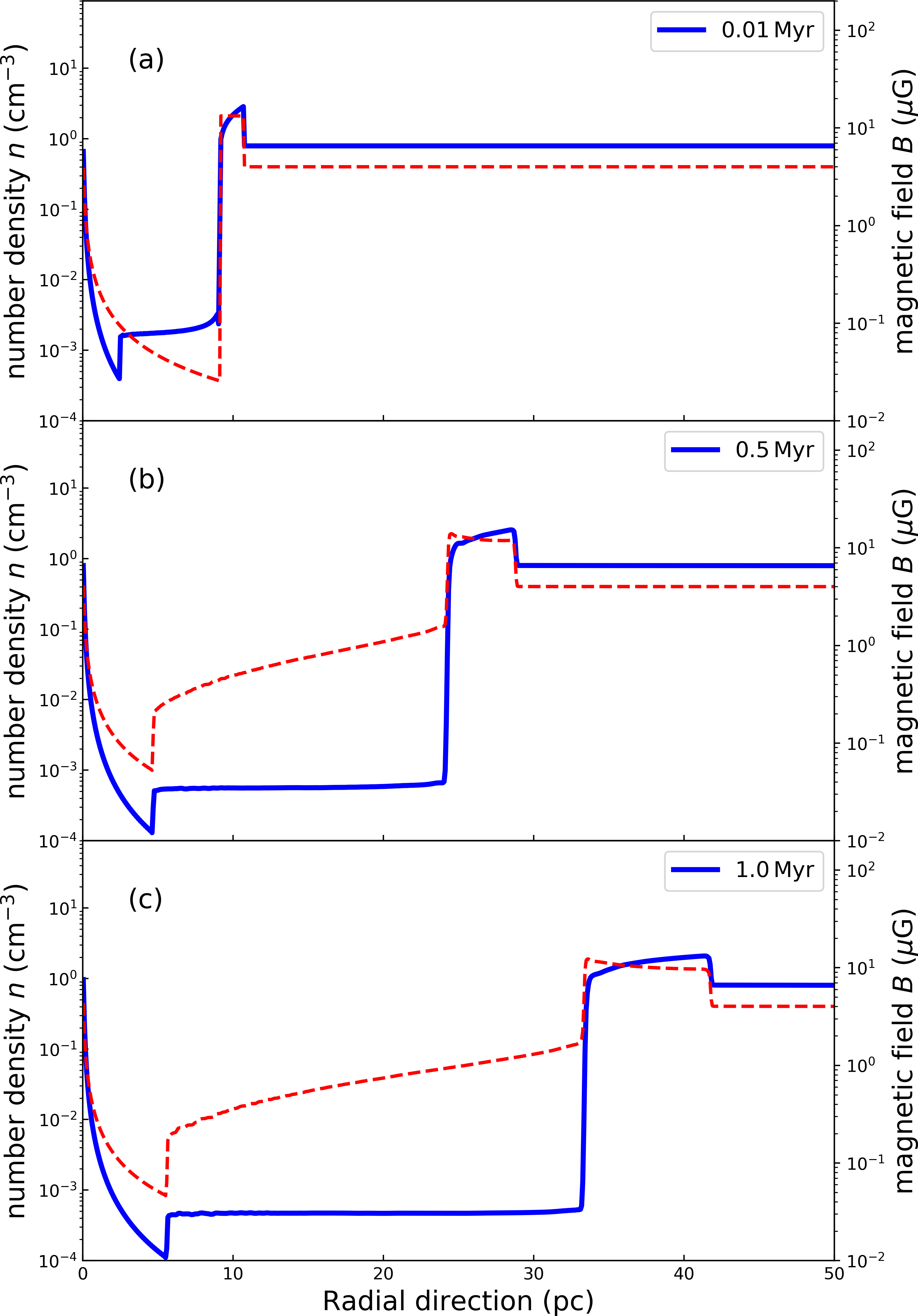}
        \caption{
        Number density (solid blue line, in $\rm cm^{-3}$) and magnetic field 
        strength (dashed red line, in $\mu \rm B$) in our simulation of the 
        stellar wind bubble of a $60\, \rm M_\odot$ star, for different times 
        of its evolution, from $0.01\, \rm Myr$ (a) to $1.0\, \rm Myr$ (c). 
        }
        \label{fig:bubble}  
\end{figure}

\begin{figure}
        \centering
        \includegraphics[width=0.42\textwidth]{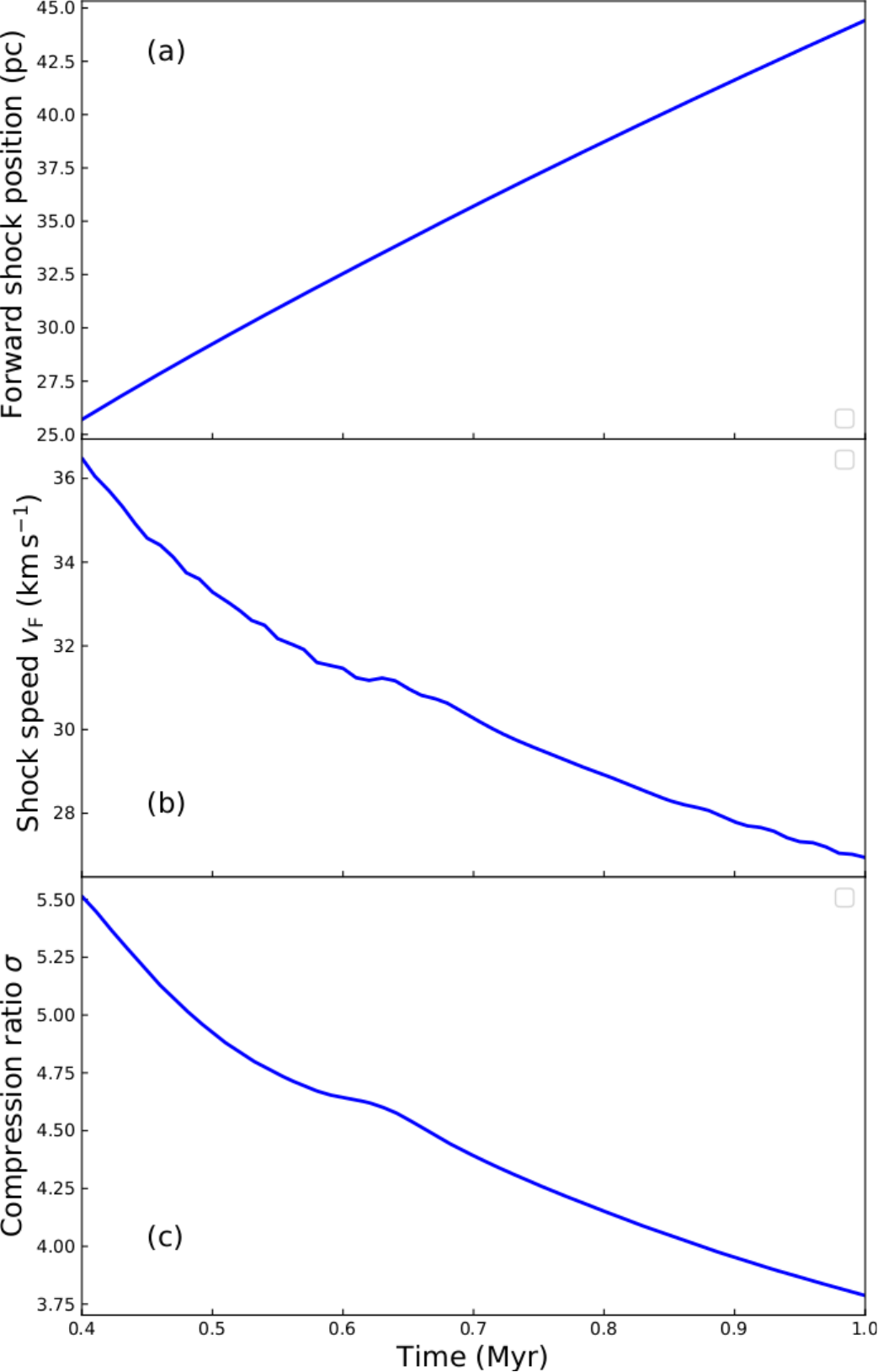}
        \caption{
        Properties of the forward shock of the stellar wind bubble of a 
        $60\, \rm M_\odot$ star. The panels show the shock position 
        $R_{\rm F}$ (in $\rm pc$), the shock speed $v_{\rm F}$ and 
        the compression ratio of the gas at the forward shock $\sigma$ 
        in the time interval between $0.4\, \rm Myr$ and $1.0\, \rm Myr$. 
        }
        \label{fig:bubble_FSH_properties}  
\end{figure}

\section{Results}
\label{sect:results}

We hereby present the 1D MHD simulations of the wind-ISM interaction of a 
$60\, \rm M_\odot$ massive star, together with the cosmic-ray distribution 
from the particle acceleration calculation and the bubble's emission spectra.

\subsection{MHD wind bubble of magnetized very massive star}
\label{sect:results_bubble}

In Fig.~\ref{fig:bubble} we show the number density field (in $\rm cm^{-3}$) 
and the magnetic field structure (in $\mu\, \rm G$) in the stellar wind bubble 
produced by wind-ISM interaction around a $60\, \rm M_\odot$ massive star. 
The initial conditions at the moment of the mapping of the initial hydrodynamical 
wind-ISM calculation as a MHD simulation on a larger grid. The density field 
already exhibits the typical structure of a wind bubble 
as described in~\citet{weaver_apj_218_1977}, with a reverse shock, a 
contact discontinuity and the forward shock solid blue line). 
The freely-expanding stellar wind characterised by a number density 
decreasing $\propto 1/r^{2}$ from the origin of the domain to the termination 
shock. A region of low-density shocked wind establishes itself between the 
termination shock and the contact discontinuity, while a dense region of 
shocked ISM material forms between the contact discontinuity and the forward 
shock, respectively (Fig.~\ref{fig:bubble}a). 
The reconstructed magnetic field consists in a Parker wind 
in the section between the origin of the domain and the contact discontinuity. 
The compressed magnetic field is imposed in the dense region of ISM gas, 
and the ISM is filled with the constant field $B_{\rm ISM}$. 
As the central star evolves through the main-sequence phase of its 
life and blows its strong wind, corresponding to a mass-loss rate 
$\dot{M}\approx 10^{-6}\, \rm M_\odot\, \rm yr^{-1}$ and wind velocity 
$v_{\rm w}\approx 2000\, \rm km\, \rm s^{-1}$. Under the effect of the 
momentum injected into the circumstellar medium, the ram pressure of 
the wind pushes the structured bubbles to larger radial distances in 
the ISM (Fig.~\ref{fig:bubble}b). The toroidal component of the magnetic 
field is compressed at the termination shock, and, as it governs the 
total magnetic field since the radial field decreases $\propto 1/r^{2}$ 
according to the stellar wind density. It then increases as a function 
of the radius from the star $r$ and is it proportional to the density 
jump in the region of shocked ISM. 
This magnetic field structure is conserved throughout the whole 
simulation (Fig.~\ref{fig:bubble}c).

\begin{figure*}
        \centering
        \includegraphics[width=1.0\textwidth]{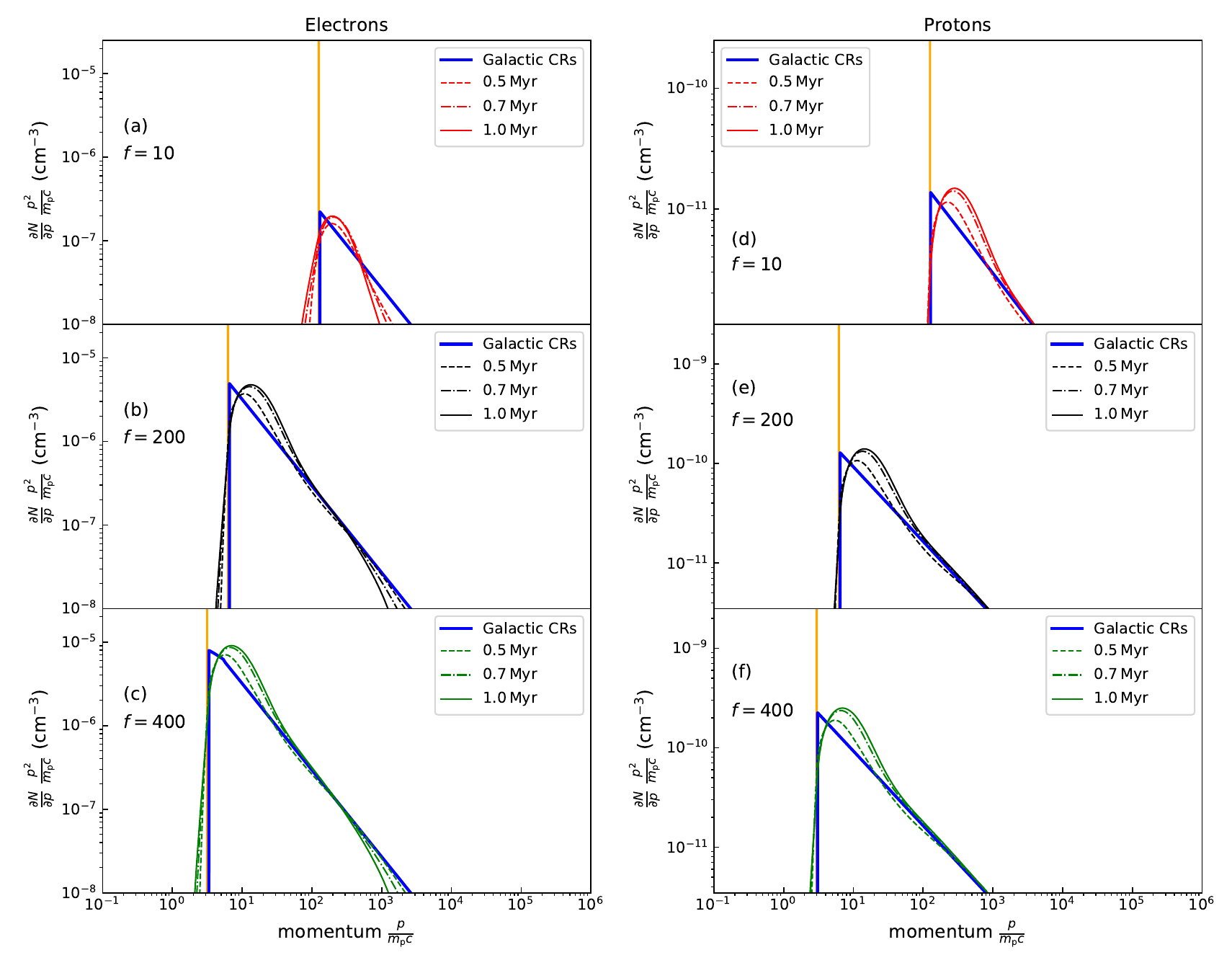}
        \caption{
        Distribution of the non-thermal particles as a function of 
        momentum (electrons, left panels), and (protons, right panels) 
        as a function of the adopted diffusion coefficient in the 
        ambient medium, controlled by the parameter $f$. 
        The number density of non-thermal particles (in $\rm cm^{-3}$) 
        is shown as a function of the normalized momentum, for 
        several time instances, $0.5\, \rm Myr$ (dotted lines), 
        $0.7\, \rm Myr$ (dashed lines) and $1.0\, \rm Myr$ (solid line). 
        The blue lines are the initial Galactic cosmic-ray background
        and the orange vertical line is the momentum threshold. 
        }
        \label{fig:part_dist_mom}  
\end{figure*}

Fig.~\ref{fig:bubble_FSH_properties} plots the properties of the expanding 
forward shock of the stellar wind bubble over the early $1\, \rm Myr$ of the 
evolution of the main-sequence $60\, \rm M_\odot$. 
At time $0.4\, \rm Myr$, the position of the forward shock is $\approx 26\, \rm pc$ 
from the star, and this distance quasi-linearly increases as a function of time 
up to times $>1\, \rm Myr$, as the volume of the wind bubble augments 
(Fig.~\ref{fig:bubble_FSH_properties}a). 
The shock speed, i.e. the velocity at which the forward shock goes through 
the local ambient medium monotonically decreases from 
$v_{\rm F}\approx 38\, \rm km\, \rm s^{-1}$ at time $0.4\, \rm Myr$ to 
$v_{\rm F}\approx 27\, \rm km\, \rm s^{-1}$ at time $1.0\, \rm Myr$, 
respectively (Fig.~\ref{fig:bubble_FSH_properties}b). 
Similarly, the value compression ratio of the forward shock decreases 
from $\sigma\approx 5.5$ to $\sigma\approx 3.8$ over the same time 
interval (Fig.~\ref{fig:bubble_FSH_properties}c). At later times, 
the expansion of the wind bubble slows down, and the forward shock 
becomes weaker as the bubble reaches its final size~\citep{meyer_mnras_493_2020}. 
We start the post-processing of the MHD wind bubble with the particle 
acceleration code at $0.4\, \rm Myr$, because at earlier times, the number 
of grid zones resolving the layer of shocked ISM between the contact 
discontinuity and the forward shock, is not sufficient to permit the shock 
finder algorithm to properly track the forward shock. 
The cosmic ray calculation is continued up to the moment the compression 
ratio at the forward shock is $\sigma \le 4$, at times $\approx 1\, \rm Myr$.

\subsection{Non-thermal particle distribution}
\label{sect:results_part}

In Fig.~\ref{fig:part_dist_mom} we display the distribution 
of non-thermal particles for electrons (left) and protons (right) as a 
function of the normalized momentum, for the different diffusion coefficient 
models, differing by the parameter $f$, see Eq.~\ref{eq:coeff_diff}. 
The number density of non-thermal particles (in $\rm cm^{-3}$) is shown 
for several time instances, $0.5\, \rm Myr$ (dotted lines), $0.7\, \rm Myr$ 
(dashed lines) and $1.0\, \rm Myr$ (solid line). The blue line is the 
initial Galactic cosmic-ray background that we assume, and which is cut-off 
at lower momentum for a given threshold energy. 
The energy threshold (orange vertical lines) is the energy under which 
no acceleration is possible, and it depends on the adopted diffusion 
model, i.e. on the parameter $f$. 

As the system evolves, a modification of the non-thermal particles distribution 
with respect to the initial Galactic cosmic-ray spectrum happens. 
An excess of non-thermal electrons above the initial population is produced 
for a diffusion model $f=10$, together with a shift of its maximum momentum to lower values,  
because of the efficient synchrotron cooling at work therein 
(Fig.~\ref{fig:part_dist_mom}a). 
A similar excess of particles is visible in the proton distribution at 
momenta larger than the initial cut-off of the Galactic background cosmic-ray 
spectrum, which does not have its maximum momentum shifted, since no 
efficient cooling is at work regarding this particle specie (Fig.~\ref{fig:part_dist_mom}d). 
The same trend can be seen in the evolution of the non-thermal electron 
and proton distribution calculated with different diffusion laws, i.e. 
a larger B\" ohm-like diffusion factor and a smaller momentum cut-off 
of the initial Galactic spectrum, providing more injected pre-existing 
particle to re-accelerate, because of the smaller threshold energy 
(Fig.~\ref{fig:part_dist_mom}b,e,c,f).

\begin{figure}
        \centering
        \includegraphics[width=0.49\textwidth]{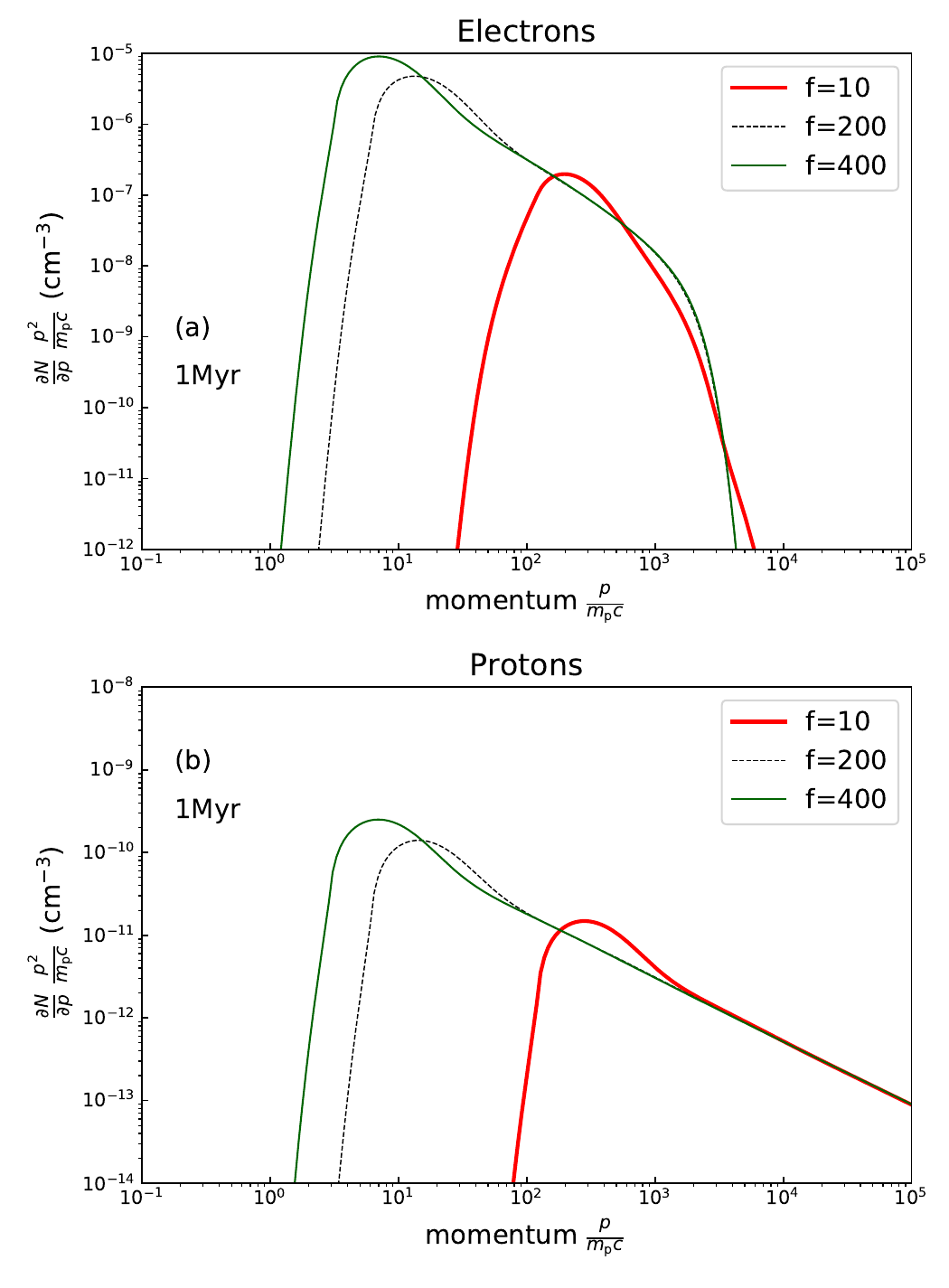}
        \caption{
        Distribution of the non-thermal particles as a function of 
        momentum (electrons, top panel; protons, bottom), 
        as a function of the adopted diffusion coefficient in 
        the ambient medium, for $f=10$ (thick solid red line), 
        $f=200$ (dotted black line) and $f=400$ (thin solid 
        dark green line). 
        The number density of non-thermal particles (in $\rm cm^{-3}$) 
        is shown as a function of the normalized momentum, at a 
        time instance corresponding to the end of the simulation 
        at time $1.0\, \rm Myr$. 
        }
        \label{fig:distribution_comparison}  
\end{figure}

Fig.~\ref{fig:distribution_comparison} compares the number density distribution of 
non-thermal particles at the end of the simulations ($1\, \rm Myr$) for 
different diffusion models ($10\le f \le 400$). 
The accelerated particles of the model with $f=10$ reach a \textcolor{black}{higher maximum 
energy} than the models with $f=200$ and $f=400$, respectively, because 
their energy threshold is higher, and their acceleration timescale 
\begin{equation}
	t_{\rm acc} = f \frac{ D_{\rm B} }{ v }, 
\end{equation}
is shorter as a result of the smaller diffusion coefficient. Hence, a 
more important accumulation of non-thermal particles with $f=10$ than 
with $f=200$ or with $f=400$ happens. 
The maximum energy $E_{\rm max}\propto v^{2}Bt$ reached by the non-thermal electrons 
is similar in all models, regardless of the parameter $f$ because of the losses 
by synchrotron emission that \textcolor{black}{are} independent of the diffusion coefficient, and 
\textcolor{black}{cool} them to the same energy (Fig.~\ref{fig:distribution_comparison}a). 
\textcolor{black}{
On the other hand, the distribution of non-thermal protons do not exhibit a 
deficit of particles with respect to the initial value of the Galactic 
cosmic-ray background, as they do not experience cooling mechanisms, see  
Fig.~\ref{fig:distribution_comparison}b. 
}

\begin{figure}
        \centering
        \includegraphics[width=0.49\textwidth]{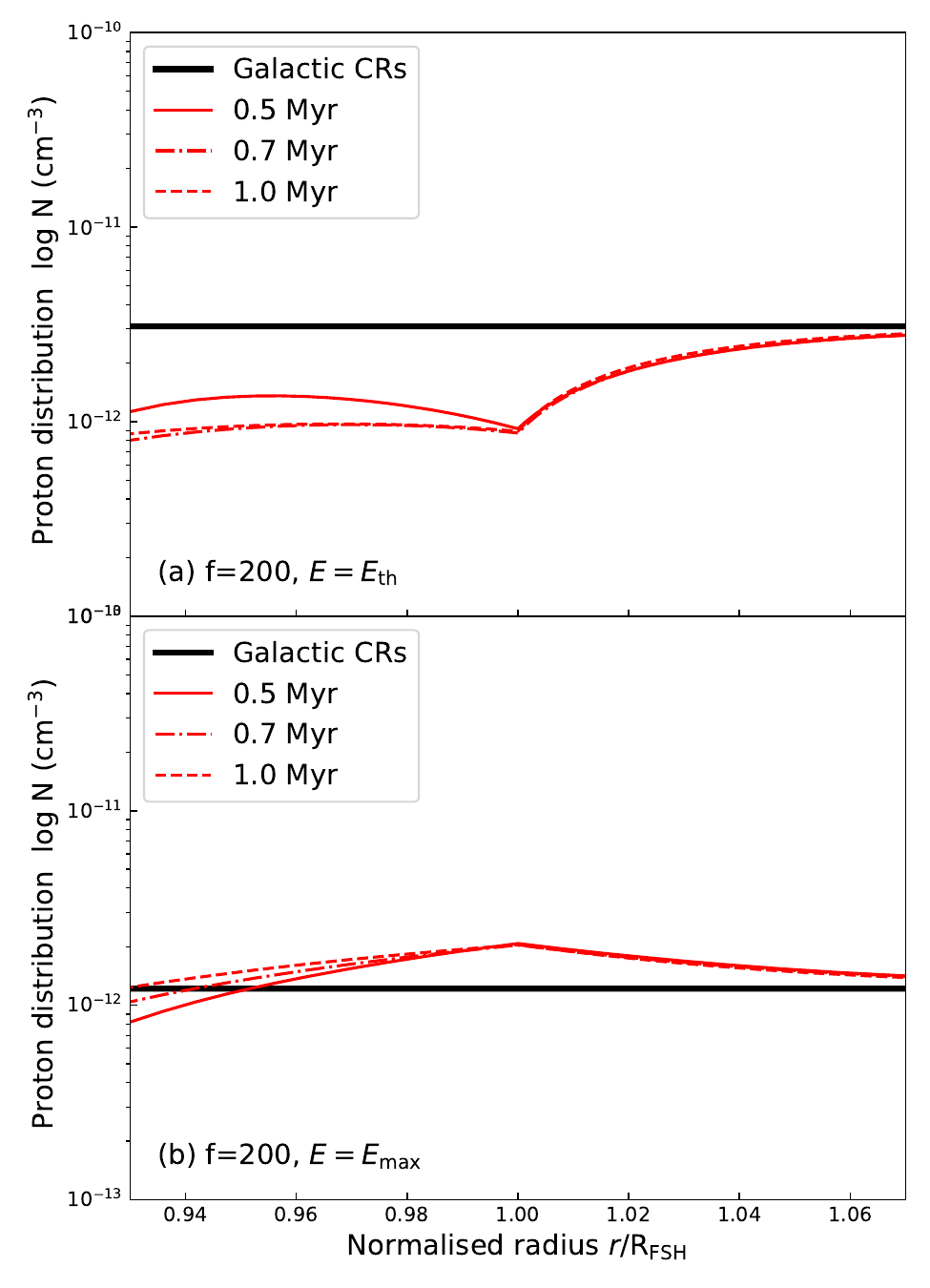}
        \caption{
        \textcolor{black}{
        Evolution of the distribution of non-thermal protons 
        (in $\rm cm^{-3}$) for the threshold (top) and maximum 
        (bottom) energies, at selected times 
        of the wind bubble evolution, in the diffusion 
        model with $f=200$. The black line is the Galactic 
        cosmic ray background. 
        }
        }
        \label{fig:distribution_evolution}  
\end{figure}

\begin{figure*}
        \centering
        \includegraphics[width=1.0\textwidth]{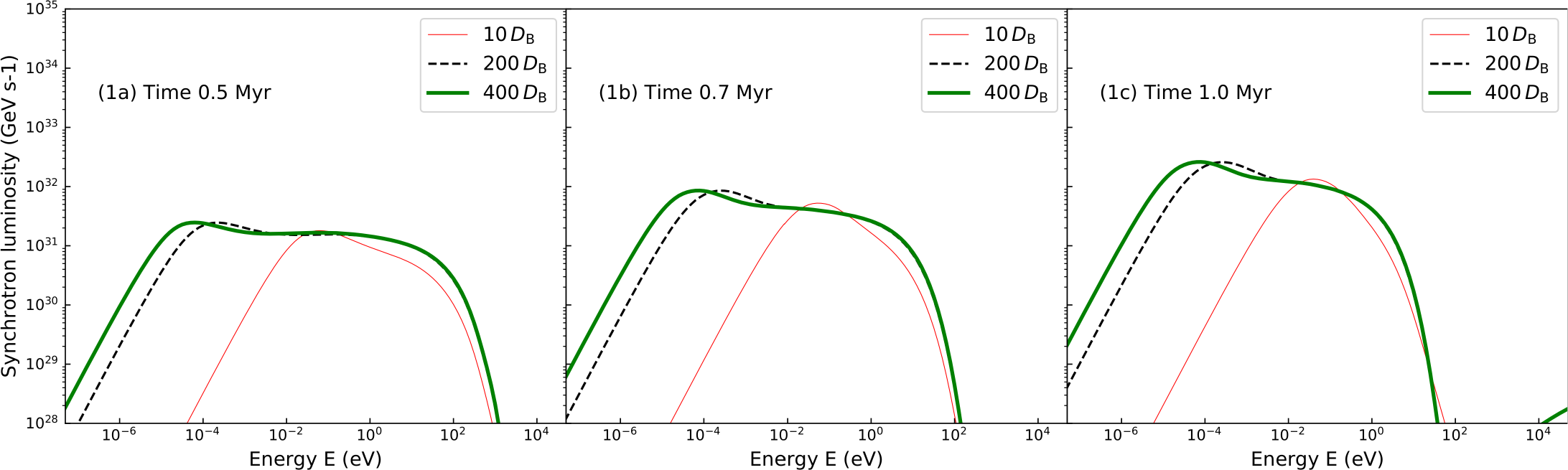} 
        \includegraphics[width=1.0\textwidth]{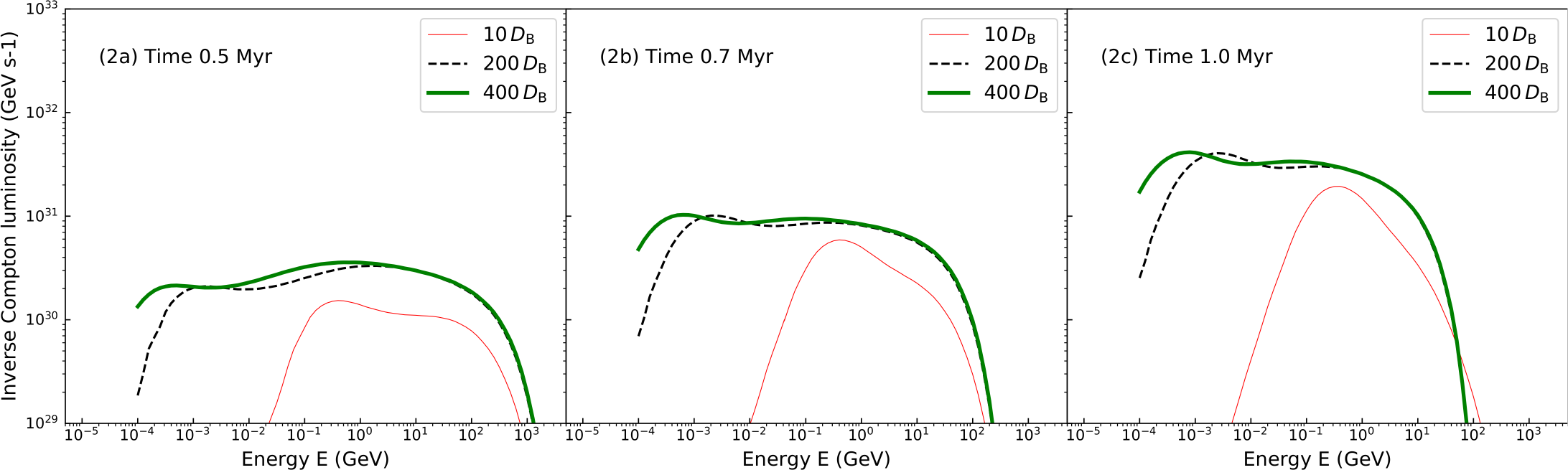} 
        \includegraphics[width=1.0\textwidth]{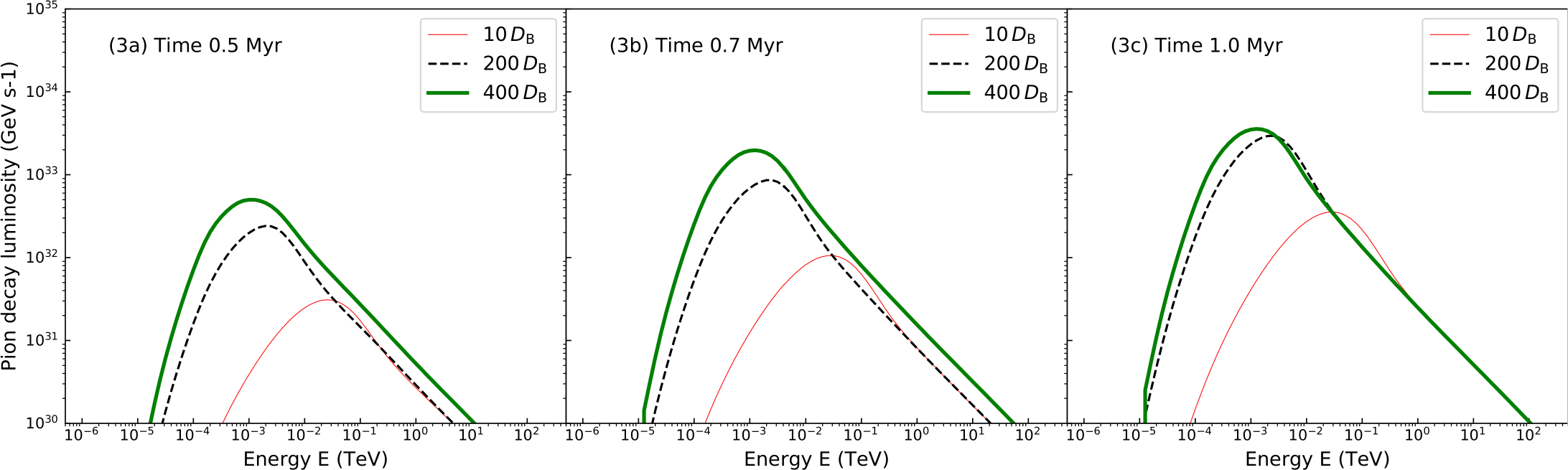}         
        \includegraphics[width=1.0\textwidth]{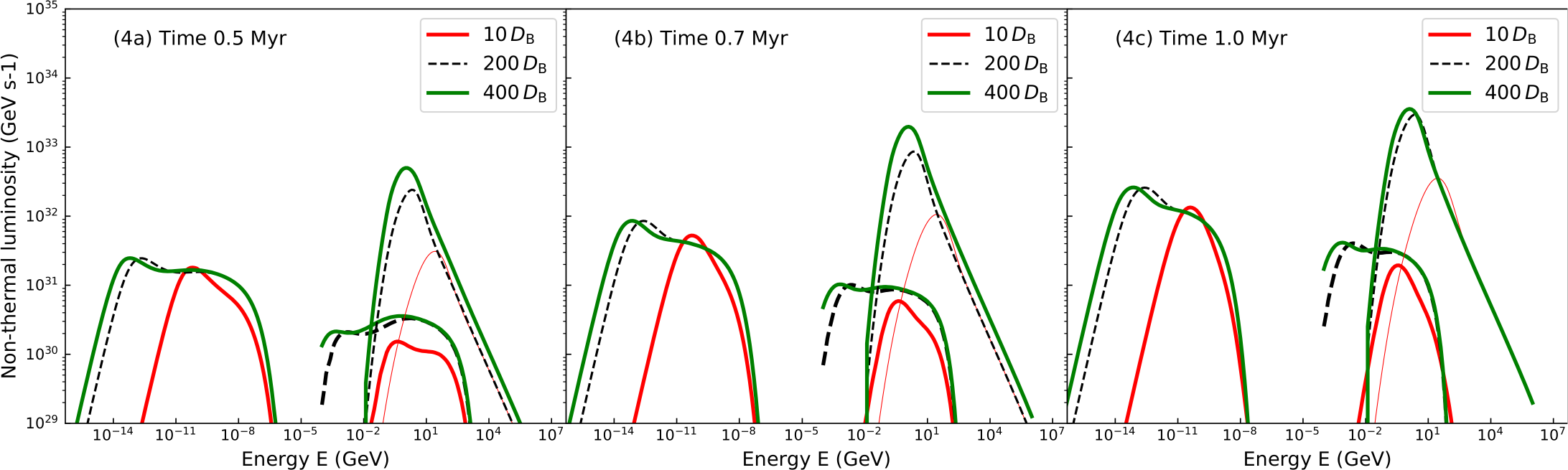} 
        \caption{
        Time-sequence evolution of the  non-thermal emission 
        spectra of the forward shock of the stellar wind bubble created 
        by a $60\, \rm M_\odot$ star during the early $1\, \rm Myr$ 
        of its main-sequence phase. 
        The luminosities are plotted as a function of the energy   
        at times $0.5\, \rm Myr$ (left), $0.7\, \rm Myr$ (middle) 
        and $1.0\, \rm Myr$ (right). 
        The emission is displayed for the synchrotron (SY, top panels), 
        inverse Compton (IC, second line of panels), $\gamma$-rays by $\pi^{o}$ 
        decay mechanisms (PD, third line of panels) and for all three 
        processes (bottom panels). 
        }
        \label{fig:sync_em_spectra}  
\end{figure*}

To better understand the acceleration process, one can have a closer look at 
the evolution of the distribution of non-thermal protons in the vicinity of 
the forward shock the stellar wind bubble, for given selected energies, 
as a function of time. 
Fig. \ref{fig:distribution_evolution} plots the radial particle distribution in 
the calculation for the diffusion model with $f=200$, for energies $E_{\rm th}$ 
and $E_{\rm max}$ throughout the simulation, for several time instances. The 
spatial coordinate is shown as normalised to the radius of the forward shock. 
The black line corresponds to the values of the pre-existing 
Galactic non-thermal bath. 
In panel Fig. \ref{fig:distribution_evolution}b, there is a peak in the 
particle distribution at the forward shock because of the confinement of 
particles due to the re-acceleration of Galactic particles, followed 
by an exponential decrease behind the forward shock. 
The slope of the radial distribution behind the shock increases since the 
magnetic field in the post-shock region decreases at the forward shock 
of the expanding wind bubble. The B\" ohm-like diffusion 
coefficient goes as $E/B$ (see solid and dashed lines of 
Fig. \ref{fig:distribution_evolution}b). 
In panel Fig. \ref{fig:distribution_evolution}a, the acceleration of particles 
at the threshold energy $E_{\rm th}$ induces a decrease of 
protons at the forward shock. There are no 
more particles with energy lower than $E_{\rm th}$ to fill 
the void: the  B\" ohm-like mechanism is energy-dependent, therefore the lower 
the energy and the longer the timsecale to diffuse such protons, hence the 
re-filling of this void is only possible by replenishing 
it with lower-energy cosmic 
rays brought by spatial advection towards the forward shock.

\subsection{Emission spectra}
\label{sect:obs}

Fig.~\ref{fig:sync_em_spectra} plots the evolution of synchrotron emission of the forward shock of the circumstellar stellar wind bubble forming around the massive star, at selected time instances during the early $1\, \rm Myr$ of its main-sequence evolution. The colors distinguish between several adopted Boehm-based diffusion models, and the several lines of panels distinguish between the emission mechanisms, i.e., synchrotron (top panels), inverse Compton scattering with cosmic microwave background photons (second line of panels), $\pi^{o}$ decays (third line of panels), and all processes together (bottom panels).
The emission flux $F$ is calculated as in~\citet{das_aa_661a_2022} at the forward shock of the wind bubble, where the particles are re-accelerated. It is converted into luminosities by integrating the flux over a volume. The luminosity represents a shell with an inner radius at the contact discontinuity and an outer radius extending $0.5\, \rm pc$ beyond the forward shock.

The fluxes of the various components of the emission spectrum increase as a 
function of time, which translates the continuous re-acceleration of pre-existing 
non-thermal particles at the forward shock of the bubble, see differences between 
the left-hand and right-hand columns of panels. The fluxes are more important 
for the synchrotron and the hadronic emission mechanism. 
The effects of cooling diminish the maximum energy of the synchrotron 
flux distribution (Fig.~\ref{fig:sync_em_spectra},1a-1c), while 
the maximum energy reached by the photon generated via inverse Compton or 
by hadronic processes is constant over time and only differs according to 
the used diffusion model (Fig.~\ref{fig:sync_em_spectra},2a-2c,,3a-3c). 
The differences caused by the adopted diffusion models show that, 
in large part, the Boehm-like coefficient corresponds 
to a higher emission flux.

The emission flux by synchrotron radiation is in the $10^{-6}$-$10^{2}\, \rm eV$ 
with a peak at about $10^{-4}\, \rm eV$ and which maximum increases, for the several 
efficient diffusion models, by a factor of 5 during the $1\, \rm Myr$ of the 
bubble's life. 
Similarly, the emission flux by the inverse Compton mechanism lies within the 
$10^{-4}$-$10^{2}\, \rm GeV$ energy band. It peaks at about $10^{-3}\, \rm GeV$ 
but displays a plateau up to about $1\, \rm GeV$, and its maximum flux increases  
by a factor of 6.6 over the early main-sequence of the star. 
The emission flux by $\pi^{o}$ decays is in the $10^{-5}$-$10^{1}\, \rm TeV$ energy 
band, peaks at about $10^{-3}\, \rm TeV$ and its maximum increases by a factor of 
5 during the simulation (Fig.~\ref{fig:sync_em_spectra}). 
The forward shock of the expanding wind bubble is therefore able to increase the 
non-thermal emission by a factor of 5 compare to the local cosmic ray background. 
The hadronic flux is more important than that by inverse Compton, which is itself 
larger than the synchrotron emission flux. The important hadronic flux comes from 
the cut-off energy of the Galactic cosmic-ray spectrum that is smaller for protons 
than for the electrons responsible for the emission by synchrotron and inverse 
Compton mechanisms.


\section{Discussion}
\label{sect:discussion}

This section presents the limitations of our model. 
Finally, we compare our findings with observational data and
\textcolor{black}{discuss} our results in the context of massive stars and 
other \textcolor{black}{observations}.

\subsection{Caveats of the model}
\label{sect:caveat}

%
First, although we time-dependently interpolate the stellar surface 
properties of its central massive star, the simulation that we perform for 
its circumstellar medium are one-dimensional, see also~\citet{weaver_apj_218_1977, 
garciasegura_1996_aa_305f,zhekov_na_3_1998,dwarkadas_apj_630_2005,dwarkadas_apj_667_2007}. 
\textcolor{black}{
This study is the first work presenting a 1D MHD model of stellar wind bubble 
including all three components of the velocity and magnetic fields, which is a 
step forward compare to the simulations of~\citet{weaver_apj_218_1977,
dwarkadas_apj_630_2005,dwarkadas_apj_667_2007}. Nevertheless, our adopted 1D 
computing strategy still suffers from an evident loss of realism compare to 
multi-dimensional simulations~\citep{freyer_apj_594_2003,freyer_apj_638_2006}. 
}
Especially, the development of instabilities at the contact 
discontinuity~\citep{meyer_mnras_493_2020}, but also the turbulence of 
the gas in the region of shocked wind~\citep{dwarkadas_apj_667_2007} 
\textcolor{black}{which change} the properties in the layer of shocked 
ISM, are not included in our 1D MHD models.

A number of physical processes at work are also neglected. The most 
prominent of all is heat conduction, affecting both the 
\textcolor{black}{thermal transfer} 
and the properties at the termination shock and at the 
forward shock of the wind bubble. This could potentially change the 
compression ratio of the shocks, and, consequently, the efficiency 
of the acceleration of electrons and protons at this emplacement. 
%
%
The ambient medium hosting the central massive star is 
setup up in its most simple description: a uniform region with properties 
corresponding to the warm phase of the ISM ($n_{\rm ISM}=0.79\, \rm cm^{-3}$ 
and $T\approx 8000\, \rm K$). 
A more realistic ambient medium, e.g. accounting for the native 
turbulence of the gas~\citep{rogers_mnras_431_2013,mackey_sept_2014}, its intrinsic 
clumpiness~\citep{wareing_mnras_470_2017,pittard_mnras_488_2019}, or the 
large-scale magnetic field of the ISM, see study of~\citet{vanmarle_584_aa_2015}.

\textcolor{black}{
A few more points can be mentioned. 
In particular, the adiabatic compression of \textcolor{black}{cosmic 
rays} in the post-shock region 
of radiative shocks, resulting from density waves travelling with high-density 
regions of the ISM, is a mechanism which could matter in the context of the 
circumstellar wind bubble~\citep{blandford_apj_260_1982}. 
The pre-existing population of Galactic comic rays would then be increased  
by a factor proportional to the compression ratio of the forward 
shock~\citep{uchiyama_apj_723_2010}, increasing the local population 
of non-thermal particles available for re-acceleration. 
This is particularly important during the early expansion phase of the bubble, 
and could magnify the hadronic emission of the stellar surroundings of massive 
stars, as it is the case in some supernova remnants~\citep{tang_apj_784_2014,
cardillo_aa_595_2016,tang_mnras_482_2019,sushch_mnras_521_2023}. 
This strengthens the results presented in this paper. 
}

\textcolor{black}{ 
Secondly, the target photon field for the inverse Compton emission 
in our calculation does not include thermal infrared photons produced 
by dust trapped in the circumstellar medium and heated by the starlight.  
This has been show to matter in the context of stellar wind 
bow shocks around OB stars~\citep{valle_ApJ_864_2018}. 
We produce our emission spectra at times about $0.5\, \rm Myr$ after the onset 
of the main-sequence phase, when the bubble is extended and its dusty ISM gas has 
already reached distances $\ge 20\, \rm pc$ (Fig. 3b). Since the starlight 
infrared flux by unit solid angle diminishes by the $1/r^2$, with $r$ the 
distance to the star, it is reasonably acceptable to neglect it in the present 
work than in~\citet{valle_ApJ_864_2018}. A more detailed study of that problem 
would be highly desirable in the future. 
}

\begin{figure}
        \centering
        \includegraphics[width=0.5\textwidth]{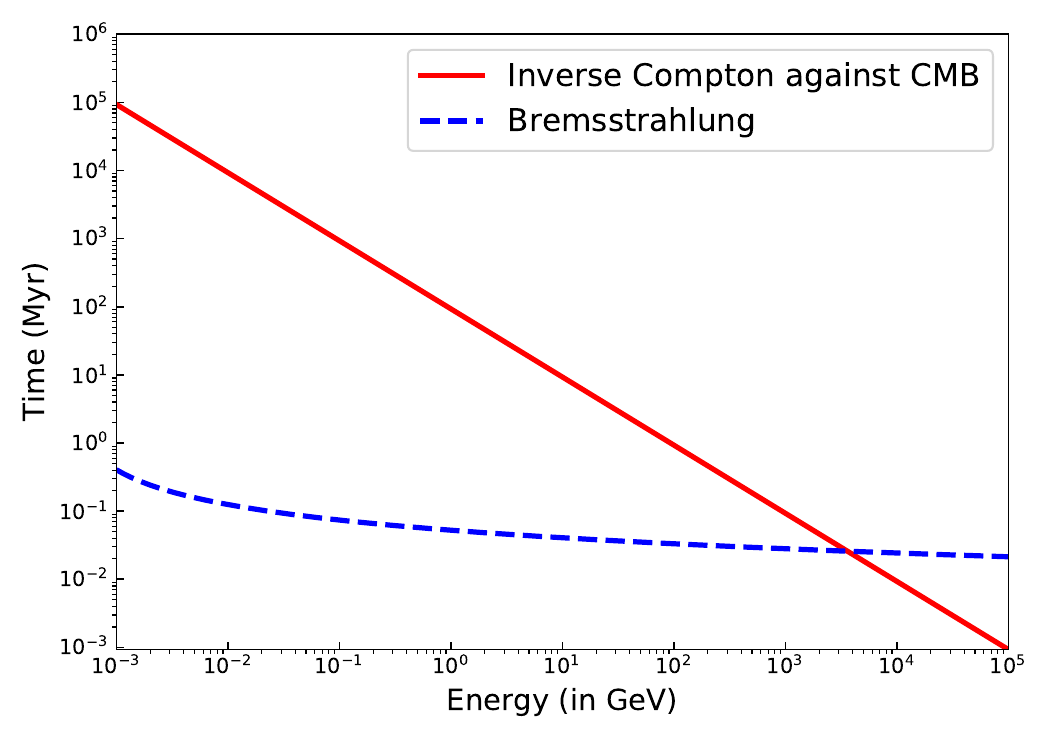} 
        \caption{
        \textcolor{black}{
        Comparison between the relativistic Bremsstrahlung 
        timescale (solid red line) and 
inverse Compton  scattering of CMB (cosmological microwave background) 
photons} (dashed blue line) timescale (in Myr), 
for the non-thermal emission at the forward shock of 
the stellar wind bubble around a massive star, 1 Myr 
after the onset of its central star main-sequence phase.
        }        
        \label{fig:timescales}  
\end{figure}

\textcolor{black}{
Third, our model does not account for non-thermal Bremsstrahlung, 
an emission mechanism of relativistic electrons emitting 
when their path is deviated by the local ISM electrical field of charged 
particles. This produces from hard X-rays to $\gamma$-rays emission. 
Since it affects supernova shocks propagating the ISM at densities similar 
to that of the warm phase of the ISM, this mechanism could affect the 
non-thermal X-rays to $\gamma$-rays emission spectrum of stellar wind bubbles 
of OB star. 
}
\textcolor{black}{
A back-of-the-envelope estimate of the non-thermal relativistic Bremsstrahlung 
and inverse Compton emission timescales can further educate us on the respective 
importance of these emission processes. We adopt the prescriptions 
of \citet{berezinskii_1990} for the Bremsstrahlung timescale and of 
\citet{longair_2011} for the inverse Compton electron lifetime.
Results are plotted in Fig. \ref{fig:timescales}, indicating that the 
Bremsstrahlung losses are quicker than the inverse Compton at energies 
$< 10^{4}\, \rm GeV$, while the inverse Compton electron lifetimes are 
shorter at higher energies $> 10^{4}\, \rm GeV$.
Consequently, one should, in the future, consider in more detail the 
relative importance of this emission mechanism. Its luminosity might 
nevertheless be negligible compared to the inverse Compton integrated 
flux, as suggested by the study on bow shocks from runaway massive 
stars in dense environments by \citet{delvalle_aa_563_2014}.
}

Our treatment of the particle acceleration mechanism is, in its turn, 
subject to limitations. We make use of the robust, but simple code 
{\sc ratpac} developed for the non-thermal spectra of blastwave from 
supernova remnants expanding in a pre-shaped 
medium~\citep{telezhinsky_aph_35_2012,wilhelm_aa_639_2020}. 
However, our approach neglects some features implemented into {\sc ratpac}, 
such as the amplification of the magnetic field ahead of the forward 
shock~\citep{brose_aa_593_2016,brose_aa_627_2019,2021arXiv210810773B}.  
%
%
Last, one should underline that spatial resolution plays a 
preponderant tole in particle acceleration simulations. Hence, 
both the 1D MHD stellar wind bubble~\citep{2021arXiv210714673P} 
and the shock reconstruction during the particle acceleration 
calculation would benefit a high number of cells discretizing 
the respective regions of interest, in the vicinity of the 
forward shock. Potential solutions could be either a new coordinate 
transformation as that adopted in this study (see Eq.~\ref{coord_transf}), 
or the use of so-called \textit{Particle-In-Cell} simulations 
to simulate the propagation of the forward shock~\citep{bohdan_phrvl_2021}.

\subsection{Generic comparison to other works}
\label{sect:obs}

The first investigation of the non-thermal phenomenon at the reverse shock 
of circumstellar wind bubbles is that of~\citet{casse_apj_237_1980}, which 
concluded on its possible but overall inefficiency, see 
also~\citep{voelk_apj_253_1982,webb_apj_298_1985}. 
Because the stellar magnetic field decreases abruptly as a function of distance 
to the star because of a Parker spiral, its value at the termination shock 
is rather small, and the efficiency of the corresponding particle acceleration 
smaller. \textcolor{black}{However, these} series of studies assumed a Weaver-like bubble, which 
stellar wind parameters are today known to be far too high in terms of mass-loss 
rate~\citep{brott_aa_530_2011a}. 
The acceleration of non-thermal particles in supernova remnants has been shown to 
add-up when multiple core-collapse explosions happen in OB associations, suggesting 
that, in the stellar clusters where the wind and supernova blastwave of several generations 
of high-mass stars collide, these effects should be very 
efficient~\citet{2001AstL...27..625B,bykov_ssrv_99_2001,butt_apj_677_2008}. 
Similarly, the study of~\citet{morlino_mnras_504_2021} demonstrates the efficient 
production of cosmic rays at the termination shock of superbubbles around clusters 
of massive stars, using semi-analytic estimates. 
These studies concern the cosmic-ray acceleration at the termination shock 
of circumstellar nebulae surrounding single massive stars and/or superbubbles 
forming around a stellar cluster of high-mass stars, which is different 
from our approach since we show how native cosmic rays in the ISM are 
re-accelerated by the forward shocks of stellar wind bubbles around massive stars, 
which acts as a magnifier of the local non-thermal particle distribution.

\begin{figure*}
        \centering
        \includegraphics[width=0.8\textwidth]{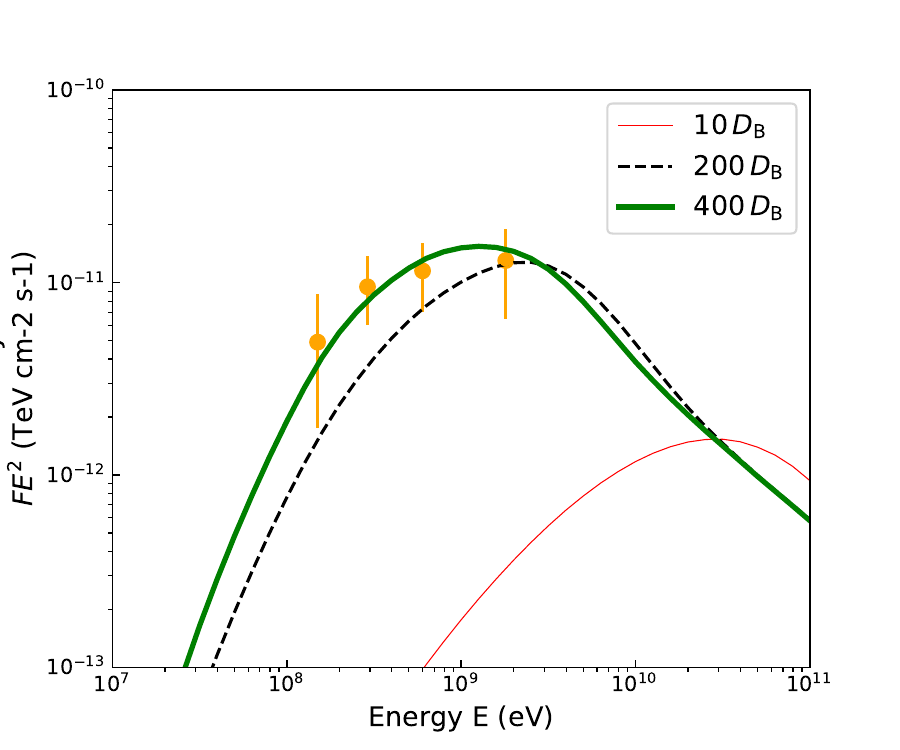}
        \caption{
        Comparison between the $\gamma$-rays spectrum of the stellar wind 
        bubble created by a $60\, \rm M_\odot$ star after $1\, \rm Myr$ 
        of main-sequence evolution, for the different adopted diffusion 
        models. 
        The plot units, frame box size and the orange dots are data from the 
        {\sc AGILE} $\gamma$-rays excess observed from the vicinity of 
        the B-type star $\kappa$ Ori and reported in fig.3-4 
        of~\citet{cardillo_aa_622_2019}. 
        }
        \label{fig:plot_comparison}  
\end{figure*}

\subsection{Comparison to the case of $\kappa$ Ori}
\label{sect:observations_kappa}

This work would not be complete without a comparison between our results and the 
observation of non-thermal emission from the surroundings of the massive  main-sequence 
B-type star $\kappa$ Ori~\citep{cardillo_aa_622_2019}. 
\textcolor{black}{
The historical star $\kappa$ Ori is located in the apparent constellation 
of Orion, also hosting the red supergiant star Betelgeuse. 
It is also a massive star, classified as a supergiant star of spectral type B0.5 Ia that 
is believed to be a variable, with moderately fluctuating 
magnitude~\citep{marchilli_aa_615_2018}. It is embedded into a large stellar wind bubble, 
produced by its own wind-ISM interaction happening during its past main-sequence phase and 
which generated a large circumstellar structure according to the theory 
of~\citet{weaver_apj_218_1977}.  
The fate of $\kappa$ Ori is that of a core-collapse supernova, which will take place 
inside of its wind-blown bubble and eventually produce a supernova remnant following 
the scenario that has been investigated in the studies 
\textcolor{black}{
of~\citet{dwarkadas_apj_630_2005,dwarkadas_apj_667_2007} 
and~\citet{das_aa_661a_2022}. 
}
Interestingly, the region of $\kappa$ Ori 
\textcolor{black}{exhibits} an hadronic emission excess which 
has been constrained to originate from the re-acceleration of cosmic rays pre-existing 
as an ISM non-thermal bath in which the circumstellar medium 
\textcolor{black}{grows, see data acquired by the 
{\sc Astro rivelatore Gamma a Immagini Leggero (AGILE)} 
high-energy facility} operating in the $\gamma$-ray energy 
band~\citep{marchilli_aa_615_2018,cardillo_aa_622_2019}. 
}

\textcolor{black}{
In Fig. \ref{fig:plot_comparison} we compare the hadronic emission spectra 
in our model with the AGILE data of the bubble around $\kappa$ Ori. The best fit 
obtained between the observational data and the present numerical results is that 
calculated with a modified B\"ohm-like diffusion model $f=400$, assuming a distance 
to the source of $260\, \rm pc$, which is of the order of that measured to 
$\kappa$ Ori~\citet{cardillo_aa_622_2019}. 
This good agreement between our simulations and the observations from the vicinity of 
$\kappa$ Ori should not erase the fact the scenario that is investigated in our 
study is slightly different from that in the data reporting cosmic-rays re-acceleration. 
Indeed, $\kappa$ Ori is into an ambient medium that is denser than the assumed 
warm phase of the Galactic plane, as its main-sequence stellar wind interacted 
with a dense cloud (see paragraph below). 
Having such consistency between numerical model and observational data within the 
context of acceleration of pre-existing cosmic rays, fully supports the fact that 
the standard B\"ohm diffusion is not appropriate to depict the non-thermal evolution 
of the bubble of $\kappa$ Ori, as demonstrated by~\citet{cardillo_galax_7_2019,
cardillo_aa_622_2019}. 
}

\textcolor{black}{
There is a degeneracy in terms of stellar surface properties (stellar 
mass-loss rate, wind velocity) and ISM density, respectively, in the shaping of 
wind-blown bubbles. A similar scaling \textcolor{black}{relation exists} 
in the shaping of stellar 
wind bow shocks around massive runaway stars, see discussion in section 3.1.3 
of~\citet{meyer_mnras_464_2017}. These elements authorise direct comparison 
between our model and the AGILE data. 
The theory of~\citet{weaver_apj_218_1977} gives the time-dependent evolution of the 
forward shock of a wind bubble around \textcolor{black}{massive stars}, 
\begin{equation}
	R_{\rm 2} = 27  n_{\rm 0}^{-1/5}   L_{\rm 36}^{1/5}  t_{\rm 6}^{3/5}, 
\end{equation}
with $n_{\rm 0}$ the ISM background number density, $L_{\rm 36}=L/10^{36}\, \rm erg\, 
\rm s^{-1}$ the stellar mechanical luminosity and $t_{\rm 6}=t/10^{6}\, \rm yr$ 
the age of the star since the onset of the stellar wind bubble. With, 
\begin{equation}
	L = \frac{1}{2} \frac{dM_{\rm w}}{dt} v_{\rm w}^2,  
\end{equation}
where $M_{\rm w}$ and $v_{\rm w}$ are the wind mass-loss rate and terminal wind
velocity, respectively, one obtains, 
\begin{equation}
	R_{\rm 2} = 27 n_{\rm 0}^{-1/5}  
    \Big(  \frac{1}{2} \frac{dM_{\rm w}}{dt} v_{\rm w}^2  \Big)^{1/5}   
    t_{\rm 6}^{3/5}, 
\end{equation}
which we can express for both the bubbles of $\kappa$ Ori and of our 
$60\, \rm M_\odot$ star, 
\begin{equation}
	R_{\rm 2}^{\kappa} = 27 n_{\rm 0, \kappa}^{-1/5}  
    \Big(  \frac{1}{2} \dot{M}_{\rm w, \kappa} v_{\rm w, \kappa}^2  \Big)^{1/5}  
    t_{\rm 6, \kappa}^{3/5}, 
\end{equation}
and, 
\begin{equation}
	R_{\rm 2}^{\rm 60\, \rm M_\odot} = 27 n_{\rm 0, \rm 60\, \rm M_\odot}^{-1/5}  
    \Big(  \frac{1}{2} \dot{M}_{\rm w, \rm 60\, \rm M_\odot} v_{\rm w, \rm 60\, \rm M_\odot}^2  \Big)^{1/5} 
    t_{\rm 6, \rm 60\, \rm M_\odot}^{3/5}, 
\end{equation}
respectively. By equalling them, 
\begin{equation}
	R_{\rm 2}^{\kappa} = R_{\rm 2}^{\rm 60\, \rm M_\odot}, 
\end{equation}
indicating that both bubbles can have similar size, 
\begin{equation}
	n_{\rm 0, \kappa}^{-1} 
    \Big( \dot{M}_{\rm w, \kappa} v_{\rm w, \kappa}^2  \Big) 
    t_{\rm 6, \kappa}^{3}
    =
    n_{\rm 0, \rm 60\, \rm M_\odot}^{-1}  
    \Big( \dot{M}_{\rm w, \rm 60\, \rm M_\odot} v_{\rm w, \rm 60\, \rm M_\odot}^2  \Big) 
    t_{\rm 6, \rm 60\, \rm M_\odot}^{3}, 
\end{equation}
one gets, 
\begin{equation}
	  \frac{ n_{\rm 0, \rm 60\, \rm M_\odot}  }{  n_{\rm 0, \kappa}    }
    \Big(
	  \frac{  t_{\rm 6, \kappa}  }{  t_{\rm 6, \rm 60\, \rm M_\odot}   }    
    \Big)^{3}
    =
    \frac{  \dot{M}_{\rm w, \rm 60\, \rm M_\odot}   }{  \dot{M}_{\rm w, \kappa}  }
    \Big(
    \frac{  v_{\rm w, \rm 60\, \rm M_\odot}    }{  v_{\rm w, \kappa}   }
    \Big)^{2}, 
\end{equation}
respectively. Using our values, $n_{\rm 0, \rm 60\, \rm M_\odot}\approx 0.79\, \rm cm^{-3}$ 
and $t_{\rm 6, \rm 60\, \rm M_\odot}=1\, \rm Myr$, as well as  
$n_{\rm 0, \kappa}\approx 30\, \rm cm^{-3}$ 
and $t_{\rm 6, \kappa}=7\, \rm Myr$~\citep{cardillo_aa_622_2019}, one finds, 
\begin{equation}
	  \frac{ n_{\rm 0, \rm 60\, \rm M_\odot}  }{  n_{\rm 0, \kappa}    }
    \Big(
	  \frac{  t_{\rm 6, \kappa}  }{  t_{\rm 6, \rm 60\, \rm M_\odot}   }    
    \Big)^{3}
    \approx 9.03. 
\end{equation}
Using the values of 
$\dot{M}_{w, 60\, \rm M_\odot} = 2 \times 10^{-6}\, \rm M_\odot\, \rm yr^{-1}$ and $v_{w, 60\, \rm M_\odot} = 2020\, \rm km\, \rm s^{-1}$ 
that we use for the main-sequence phase of the massive 
star~\citep{groh_aa564_2014,meyer_mnras_493_2020} 
and similarly taking 
$\dot{M}_{\rm w, \kappa} = 2 \times 10^{-7}\, \rm M_\odot\, \rm yr^{-1}$ and $v_{\rm w, \kappa} = 2000\, \rm km\, \rm s^{-1}$ 
as in~\citet{Lennon_aa_252_1991,cardillo_aa_622_2019}, \textcolor{black}{one finds}, 
\begin{equation}
    \frac{  \dot{M}_{\rm w, \rm 60\, \rm M_\odot}   }{  \dot{M}_{\rm w, \kappa}  }
    \Big(
    \frac{  v_{\rm w, \rm 60\, \rm M_\odot}    }{  v_{\rm w, \kappa}   }
    \Big)^{2} 
    \approx 10.2.  
\end{equation}
Hence, our adopted prescription for the wind power and the modelled time window of the 
star's main-sequence phase are consistent with the current knowledge that we have 
regarding $\kappa$ Ori, in terms of dimensions of its circumstellar wind bubble. 
}

Since many uncertainties persist regarding the values of the mass-loss rate and 
the wind velocity of $\kappa$ Ori, it could be blowing a stronger wind and still 
have the same size, as long as the compression ratio at the forward shock is more 
important. 
In the context of a radiative shock, the compression ratio can reach 
high values $> 10$ as suggested in~\citet{cardillo_aa_622_2019}; 
however, the magnitude of the ISM magnetic field is also 
unconstrained, which could lead to diminishing effects at the forward 
shock~\citep{meyer_mnras_464_2017,meyer_mnras_502_2021}, thereby 
altering the re-acceleration mechanism of pre-existing particles. 
Further, more detailed simulations are necessary to fine-tune these 
particular situations and better understand the excess $\gamma$-ray 
emission from massive star wind bubbles.

\subsection{The X-rays surroundings of $\kappa$ Ori}
\label{sect:radio}

\textcolor{black}{
The overall surroundings of $\kappa$ Ori is a complex environment that is accessible 
for study through multi-wavelength observations with high-angular resolution data 
acquisition techniques. In particular, the study by ~\citet{pillitteri_apj_820_2016} 
presents data obtained with the {\it X-ray Multi-Mirror (XMM)-Newton} space-based 
observatory, operating in the soft X-ray waveband, of the immediate vicinity of $\kappa$ 
Ori. This work proposes that a large and dense shell of swept-up material has formed 
around the central massive star. The X-ray fluxes are estimated to be in the range of 
$0.8-7.01\, \times 10^{-14}\, \rm erg\, 
\rm s^{-1}\, \rm cm^{-2}\, \approx 6.24\, \times 10^{-12}$$-$$4.43\, \times 10^{-11}\, 
\rm GeV\, \rm s^{-1}\, \rm cm^{-2}$, and are interpreted as originating from the nearby 
hundreds of low-mass protostars that have formed in a dusty ring around $\kappa$ Ori. 
These protostars were found to obscure their reflected infrared starlight emission on 
the dust particles in past near-infrared observational surveys.
}
\textcolor{black}{
Since the X-ray fluxes of the $\kappa$ Ori region are much larger than those that 
we predict, we conclude that our modeling is consistent with the observations, 
suggesting that the real data may consist of two components: one from the young 
stellar objects of the dusty ring and another from the re-accelerated cosmic rays, 
the latter being several orders of magnitude fainter than the former. 
This additional element supports the conclusions of ~\citet{cardillo_aa_622_2019}, 
stating that the circumstellar medium of $\kappa$ Ori is a cosmic-ray re-accelerator.
}

\subsection{The radio synchrotron emission of the surroundings of $\kappa$ Ori}
\label{sect:radio}

\textcolor{black}{
The numerical simulations presented in this study predict non-thermal emission produced by the synchrotron mechanism (see panel 1a of Fig.~\ref{fig:sync_em_spectra}). In the energy range $10^{-12}$--$10^{-7}$ GeV, which falls within the radio waveband ($\mathrm{mm}$ and longer wavelengths), these luminosities reach $10^{31}$ GeV s$^{-1}$, corresponding to fluxes on the order of $1.0$--$1.1 \times 10^{-10}$ GeV s$^{-1}$ cm$^{-2}$, or, in other cgs units, $1.60 \times 10^{-13}$ erg s$^{-1}$ cm$^{-2}$. Since synchrotron radio emission has been detected from the Orion region, this allows for a qualitative comparison between such measurements and our predictions.
These multi-wavelength observations have, among others, been performed with the Very Large Array ({\itshape VLA}) and the Atacama Large Millimeter/submillimeter Array ({\itshape ALMA}) in the radio waveband, reporting variability in the light curves of young stellar objects~\citep{massi_aa_453_2006,forbrich_aa_477_2008,rivilla_apj_808_2015,Forbrich_apj_844_2017,2021csss.confE.234V}, providing a high-energy non-thermal counterpart to the pre-main-sequence flares affecting protostars, whose origin has been proposed within the burst mode of accretion in star formation scenarios~\citep{voroboyov_apj_650_2006,vorobyov_apj_704_2009,vorobyov_apj_719_2010,vorobyov_aa_613_2018,meyer_487_MNRAS_2019,meyer_mnras_500_2021,meyer_mnras_517_2022}.
}

\textcolor{black}{
Additionally, several other radio sources in Orion, observed with the Arecibo telescope and the VLA, have been constrained as the outcome of synchrotron emission mechanisms occurring at shocks generated by outflows from higher-mass stars, such as a Herbig-Haro object~\citep{Yusef_apj_343_1990}, the edge of the Orion-Eridanus superbubble~\citep{bracco_aa_677_2023}, or a more complex structure, e.g. consisting of relic supernova remnants and bow shock nebulae around runaway pulsars~\citep{west_apj_941_2022}.
The interpretation of the synchrotron radio emission in the Orion region as originating from particles accelerated at the edges of protostellar jets or other circumstellar shocks, like the expanding forward shock of an old supernova remnant or a stellar wind bubble, is not contradictory to the alternative scenario in which pre-existing cosmic rays would be re-accelerated, or even to a third hypothesis in which both mechanisms occur simultaneously.
Hence, our model is consistent with the presence of synchrotron radio emission in the Orion region and might contribute, at least in part, to their explanation as originating from the shocked environments of massive stars like $\kappa$ Ori.
}

\subsection{Comparison to stellar wind bow shocks}
\label{sect:observations_bowsocks}

This work is consistent with the \textcolor{black}{interpretation of}   
observational results on stellar wind bubbles and stellar 
wind bow shocks who either reported or claimed the importance of cosmic-ray 
acceleration in massive stellar surroundings, because of the acceleration of 
particles at the termination shock of circumstellar structures such as stellar 
wind bow shocks~\citep{delvalle_aa_543_2012, 
delvalle_aa_563_2014,valle_mnras_448_2015,valle_ApJ_864_2018}. 
Particularly, the bow shock EB27 around the massive OB star BD+43${\degree}$3654 
has been predicted and shown to be a non-thermal emitter~\citep{benaglia_aa_517_2010, 
delvalle_aa_550_2013,palacio_aa_617_2018,sanchezayaso_apj_861_2018,benaglia_mnras_503_2021}, 
although other searches for non-thermal emission in stellar surroundings have been 
unfruitful~\citep{schultz_aa_565_2014,toala_apj_821_2016,tola_apj_838_2017,
DeBecker_mnras_471_2017,binder_aj_157_2019}, suggesting that if the mechanisms of 
particle re-/acceleration is at work therein, it must be of lower efficiency 
than in other astrophysical sources such as AGN, pulsars winds, or supernova remnants.

\textcolor{black}{
BD+60${\degree}$2522 is an O-type massive star driving the nebula NGC 7635, 
well-known for its multi-wavelength emission, such as optical and  
infrared~\citep{moore_aj_2002}. Its stellar wind power and its space 
velocity are weaker than that of BD+43${\degree}$3654, 
however, \textcolor{black}{its ambient medium is denser than that of BD+43${\degree}$3654}, so that the 
numerical work of~\citet{green_aa_625_2019} suggested first the Bubble 
Nebula could be a potential non-thermal emitter. 
Previous observational campaigns of the XMM-Newton space-born telescope of 
the Bubble Nebula revealed neither thermal nor non-thermal X-ray emission, 
see~\citet{toala_apj_821,toala_mnras_495_2020}. 
The study of~\citet{2022arXiv220411913M} deepens this effort using original 
interferometric radio continuum data at frequencies 4–12 GHz with the 
{\sc Very Large Array}, rectified with single dish polarimetric observation 
at 4–8 GHz using the Effelsberg radio. 
The observational studies of~\citet{2022arXiv220411913M} thus revealed that a second 
stellar wind bow shock around the surroundings of the runaway massive star 
NGC 7635 also emits detectable non-thermal radiation. 
}

The nature of the non-thermal emission is determined by mapping the spectral 
index of the bow shocks and by comparing the flux of their spectral energy 
distribution with one-zone models for the acceleration/compression of wind/ISM 
non-thermal particles.  
Two different mechanisms have been invoked when considering non-thermal emission 
of stellar wind bow shocks of massive runaway stars: on the one hand, the emission 
of particles accelerated at the termination of the circumstellar medium and emitting 
in the shocked magnetic field~\citep{valle_ApJ_864_2018}, on the other hand, the 
re-acceleration of pre-existing particles of the ISM at the forward shock 
of the star's bow shocks~\citep{cardillo_aa_622_2019,2022arXiv220411913M}. 
The second process is consistent with the surroundings of BD+43${\degree}$3654, 
while unrealistic parameters at the forward shock are necessary to explain 
those of BD+60${\degree}$2522. The first process could not fit the data for 
both bow shocks. Since bow shocks are compact nebulae with strong shocks, 
the ambient medium of BD+43${\degree}$3654 has a density of $n \approx 15\, \rm cm^{-3}$, 
which is less than that measured for $\kappa$ Ori, and both wind terminal 
velocities are similar ($\approx 2020\, \rm km\, \rm s^{-1}$), 
but its mass-loss rate is larger ($\dot{M} = 9 \times 10^{-6}\, 
\rm M_\odot\, \rm yr^{-1}$), one can suggest that the radiation of 
accelerated particles 
might not be the appropriate mechanism
to explain the $\gamma$-ray excess of the wind bubble of $\kappa$ Ori. 
This supports the conclusions of $\kappa$ Ori being a cosmic 
rays re-accelerator~\citet{cardillo_aa_622_2019}.


\section{Conclusion}
\label{sect:conclusion}

In this study, we explore the possibility of shocks in a stellar wind 
bubble~\citep{weaver_apj_218_1977} forming around very massive stars 
to be the site of particle acceleration and to \textcolor{black}{generate non-thermal} 
emission~\citep{casse_apj_237_1980,voelk_apj_253_1982,webb_apj_298_1985}. 
The wind-ISM interaction shaping the circumstellar medium 
of a $60\, \rm M_\odot$~\citep{groh_aa564_2014} is modeled by means 
of 1D MHD numerical simulations using the {\sc pluto} 
code~\citep{mignone_apj_170_2007,migmone_apjs_198_2012,vaidya_apj_865_2018}, 
and further post-processed with the particle acceleration code 
{\sc ratpac}~\citep{telezhinsky_aph_35_2012,telezhinsky_aa_541_2012, 
telezhinsky_aa_552_2013}. 
The 1D MHD simulation account for the self-consistent time-dependent 
evolution of the stellar surface magnetic field, based on the Parker 
solution~\citep{parker_paj_128_1958} and on the stellar properties 
of~\citet{groh_aa564_2014}. 
The MHD structure of the wind bubble is further analyzed by a particle 
acceleration code, within the test particle approximation, which runs 
a shock-finder to catch its forward shock and calculate the 
acceleration of particles there. 
We obtain the time-dependent non-thermal particle density of accelerated 
electrons and protons at the shock, together with their radiation 
properties by synchrotron, inverse Compton and hadronic emission, 
respectively.

We show that the forward shock of a young stellar wind bubble around 
an early-type main-sequence very massive star, that is expanding into 
its local ambient medium, is able to re-accelerate pre-existing Galactic 
cosmic rays. 
The shock at the outer young bubble is sufficiently compressed to have  
a compression ratio $\ge 4$ and to accelerate the non-thermal electrons 
and protons, which have diffused from the \textcolor{black}{ISM} to the vicinity 
of the forward shock. The stellar wind bubble radiates via synchrotron, 
inverse Compton mechanisms and emits $\gamma$-ray emission by $\pi^{o}$ decay. 
The emission flux spans from the sub-eV to the TeV energy bands, with a 
dominant component of the hadronic component at the GeV energy band and 
are multiplied by a factor of 5 during the early expansion of the bubble. 
This process is at work in the initial expansion phase of the bubble, 
since the ISM magnetic field damps the shock at later times and the 
shock ceases to be radiative at later times.

\textcolor{black}{
Our results are qualitatively in accordance with the enhanced $\gamma$-ray emission at 
about $1\, \rm GeV$ measured from the surroundings of the massive star $\kappa$ Ori, 
This emission has been shown to be inconsistent with diffusive shock acceleration 
at the forward shock of its circumstellar bubble, but compatible with the re-acceleration 
of ISM cosmic rays~\citep{cardillo_aa_622_2019}. 
%
%
This phenomenon likely applies to all bubbles of static massive stars, as well as the 
bow shocks of runaway stars ~\citep{2022arXiv220411913M}. Therefore, such a process 
should not be omitted when seeking to better understand the Galactic cosmic-ray 
spectrum~\citep{cardillo_galax_7_2019}. This study needs to be extended to a broader 
scope, exploring both the stellar properties and the local conditions of the ISM. 
}


\section*{Acknowledgements}

\textcolor{black}{
The author acknowledges the anonymous referee for advice, which greatly improved the 
quality of this study, as well as I. Telezhinsky and A.~Wilhelm for their assistance 
regarding the {\sc ratpac} code.  
}
The authors acknowledge the North-German Supercomputing Alliance (HLRN) for providing HPC resources 
that have contributed to the research results reported in this paper. 
\textcolor{black}{
This work has been supported by the grant PID2021-124581OB-I00 funded by 
MCIN/AEI/10.13039/501100011033 and 2021SGR00426 of the Generalitat de Catalunya. 
This work was also supported by the Spanish program Unidad de Excelencia Mar\' ia 
de Maeztu CEX2020-001058-M. This work also supported by MCIN with funding from 
European Union NextGeneration EU (PRTR-C17.I1). 
}

\section*{Data availability}

\title[Wind bubbles as cosmic-rays re-accelerators]{
\textcolor{black}{Stellar wind bubbles of OB stars as Galactic cosmic-ray re-accelerators}
}

This research made use of the {\sc pluto} code developed at the University of Torino  
by A.~Mignone (http://plutocode.ph.unito.it/). The figures have been produced using the 
Matplotlib plotting library for the Python programming language (https://matplotlib.org/). 
The {\sc ratpac} code makes use of the {\sc fipy} library 
(https://www.ctcms.nist.gov/fipy/index.html). 
The data underlying this article will be shared on reasonable request to the corresponding 
author.


\bibliographystyle{mn2e}

\footnotesize{
\bibliography{grid}

\begin{thebibliography}{}

\bibitem[\protect\citeauthoryear{{Abdo}, {Ackermann} \& {Ajello}}{{Abdo}
  et~al.}{2010}]{abdo_apj_722_2010}
{Abdo} A.~A.,  {Ackermann} M.,    {Ajello} M.,  2010, \apj, 722, 1303

\bibitem[\protect\citeauthoryear{{Acero}}{{Acero}}{2023}]{acero_cta_2023}
{Acero} F. e.~a.,  2023, Astroparticle Physics, 150, 102850

\bibitem[\protect\citeauthoryear{{Acharyya}}{{Acharyya}}{2023}]{acharyya_mnras_523_2023}
{Acharyya} A. e.~a.,  2023, \mnras, 523, 5353

\bibitem[\protect\citeauthoryear{{Acreman}, {Stevens} \& {Harries}}{{Acreman}
  et~al.}{2016}]{acreman_mnras_456_2016}
{Acreman} D.~M.,  {Stevens} I.~R.,    {Harries} T.~J.,  2016, \mnras, 456, 136

\bibitem[\protect\citeauthoryear{{Avedisova}}{{Avedisova}}{1972}]{avedisova_saj_15_1972}
{Avedisova} V.~S.,  1972, \sovast, 15, 708

\bibitem[\protect\citeauthoryear{{Axford}, {Leer} \& {Skadron}}{{Axford}
  et~al.}{1977}]{axford_icrvc_1977}
{Axford} W.~I.,  {Leer} E.,    {Skadron} G.,  1977, in International Cosmic Ray
  Conference Vol.~11 of International Cosmic Ray Conference, {The Acceleration
  of Cosmic Rays by Shock Waves}.
p.~132

\bibitem[\protect\citeauthoryear{{Baalmann}, {Scherer}, {Kleimann}, {Fichtner},
  {Bomans} \& {Weis}}{{Baalmann} et~al.}{2021}]{baalmann_aa_650_2021}
{Baalmann} L.~R.,  {Scherer} K.,  {Kleimann} J.,  {Fichtner} H.,  {Bomans}
  D.~J.,    {Weis} K.,  2021, \aap, 650, A36

\bibitem[\protect\citeauthoryear{{Bednarek} \& {Bartosik}}{{Bednarek} \&
  {Bartosik}}{2004}]{bednarek_aa_423_2004}
{Bednarek} W.,  {Bartosik} M.,  2004, \aap, 423, 405

\bibitem[\protect\citeauthoryear{{Bell}}{{Bell}}{1978a}]{bell_mnras_182_1978a}
{Bell} A.~R.,  1978a, \mnras, 182, 147

\bibitem[\protect\citeauthoryear{{Bell}}{{Bell}}{1978b}]{bell_mnras_182_1978b}
{Bell} A.~R.,  1978b, \mnras, 182, 443

\bibitem[\protect\citeauthoryear{{Benaglia}, {del Palacio}, {Hales} \&
  {Colazo}}{{Benaglia} et~al.}{2021}]{benaglia_mnras_503_2021}
{Benaglia} P.,  {del Palacio} S.,  {Hales} C.,    {Colazo} M.~E.,  2021,
  \mnras, 503, 2514

\bibitem[\protect\citeauthoryear{{Benaglia}, {Romero}, {Mart{\'{\i}}}, {Peri}
  \& {Araudo}}{{Benaglia} et~al.}{2010}]{benaglia_aa_517_2010}
{Benaglia} P.,  {Romero} G.~E.,  {Mart{\'{\i}}} J.,  {Peri} C.~S.,    {Araudo}
  A.~T.,  2010, \aap, 517, L10

\bibitem[\protect\citeauthoryear{{Berezinskii}, {Bulanov}, {Dogiel} \&
  {Ptuskin}}{{Berezinskii} et~al.}{1990}]{berezinskii_1990}
{Berezinskii} V.~S.,  {Bulanov} S.~V.,  {Dogiel} V.~A.,    {Ptuskin} V.~S.,
  1990, {Astrophysics of cosmic rays}

\bibitem[\protect\citeauthoryear{{Binder}, {Behr} \& {Povich}}{{Binder}
  et~al.}{2019}]{binder_aj_157_2019}
{Binder} B.~A.,  {Behr} P.,    {Povich} M.~S.,  2019, \aj, 157, 176

\bibitem[\protect\citeauthoryear{{Blandford} \& {Cowie}}{{Blandford} \&
  {Cowie}}{1982}]{blandford_apj_260_1982}
{Blandford} R.~D.,  {Cowie} L.~L.,  1982, \apj, 260, 625

\bibitem[\protect\citeauthoryear{{Blandford} \& {Ostriker}}{{Blandford} \&
  {Ostriker}}{1978}]{blandford_apj_221_1978}
{Blandford} R.~D.,  {Ostriker} J.~P.,  1978, \apjl, 221, L29

\bibitem[\protect\citeauthoryear{{Blasi}}{{Blasi}}{2011}]{blasi_crpa_conf_2011}
{Blasi} P.,  2011, in {Giani} S.,  {Leroy} C.,   {Rancoita} P.~G.,  eds, Cosmic
  Rays for Particle and Astroparticle Physics {Cosmic Ray Acceleration in
  Supernova Remnants}.
pp 493--506

\bibitem[\protect\citeauthoryear{{Bohdan}, {Pohl}, {Niemiec}, {Morris},
  {Matsumoto}, {Amano}, {Hoshino} \& {Sulaiman}}{{Bohdan}
  et~al.}{2021}]{bohdan_phrvl_2021}
{Bohdan} A.,  {Pohl} M.,  {Niemiec} J.,  {Morris} P.~J.,  {Matsumoto} Y.,
  {Amano} T.,  {Hoshino} M.,    {Sulaiman} A.,  2021, \prl, 126, 095101

\bibitem[\protect\citeauthoryear{{Bracco}, {Padovani} \& {Soler}}{{Bracco}
  et~al.}{2023}]{bracco_aa_677_2023}
{Bracco} A.,  {Padovani} M.,    {Soler} J.~D.,  2023, \aap, 677, L11

\bibitem[\protect\citeauthoryear{{Brose}, {Pohl} \& {Sushch}}{{Brose}
  et~al.}{2021}]{2021arXiv210810773B}
{Brose} R.,  {Pohl} M.,    {Sushch} I.,  2021, arXiv e-prints, p.
  arXiv:2108.10773

\bibitem[\protect\citeauthoryear{{Brose}, {Sushch}, {Pohl}, {Luken},
  {Filipovi{\'c}} \& {Lin}}{{Brose} et~al.}{2019}]{brose_aa_627_2019}
{Brose} R.,  {Sushch} I.,  {Pohl} M.,  {Luken} K.~J.,  {Filipovi{\'c}} M.~D.,
   {Lin} R.,  2019, \aap, 627, A166

\bibitem[\protect\citeauthoryear{{Brose}, {Telezhinsky} \& {Pohl}}{{Brose}
  et~al.}{2016}]{brose_aa_593_2016}
{Brose} R.,  {Telezhinsky} I.,    {Pohl} M.,  2016, \aap, 593, A20

\bibitem[\protect\citeauthoryear{{Brott}, {de Mink}, {Cantiello}, {Langer}, {de
  Koter}, {Evans}, {Hunter}, {Trundle} \& {Vink}}{{Brott}
  et~al.}{2011}]{brott_aa_530_2011a}
{Brott} I.,  {de Mink} S.~E.,  {Cantiello} M.,  {Langer} N.,  {de Koter} A.,
  {Evans} C.~J.,  {Hunter} I.,  {Trundle} C.,    {Vink} J.~S.,  2011, \aap,
  530, A115

\bibitem[\protect\citeauthoryear{{Butt} \& {Bykov}}{{Butt} \&
  {Bykov}}{2008}]{butt_apj_677_2008}
{Butt} Y.~M.,  {Bykov} A.~M.,  2008, \apjl, 677, L21

\bibitem[\protect\citeauthoryear{{Bykov}}{{Bykov}}{2001}]{bykov_ssrv_99_2001}
{Bykov} A.~M.,  2001, \ssr, 99, 317

\bibitem[\protect\citeauthoryear{{Bykov} \& {Toptygin}}{{Bykov} \&
  {Toptygin}}{2001}]{2001AstL...27..625B}
{Bykov} A.~M.,  {Toptygin} I.~N.,  2001, Astronomy Letters, 27, 625

\bibitem[\protect\citeauthoryear{{Cardillo}}{{Cardillo}}{2019}]{cardillo_galax_7_2019}
{Cardillo} M.,  2019, Galaxies, 7, 49

\bibitem[\protect\citeauthoryear{{Cardillo}, {Amato} \& {Blasi}}{{Cardillo}
  et~al.}{2016}]{cardillo_aa_595_2016}
{Cardillo} M.,  {Amato} E.,    {Blasi} P.,  2016, \aap, 595, A58

\bibitem[\protect\citeauthoryear{{Cardillo}, {Marchili}, {Piano}, {Giuliani},
  {Tavani} \& {Molinari}}{{Cardillo} et~al.}{2019}]{cardillo_aa_622_2019}
{Cardillo} M.,  {Marchili} N.,  {Piano} G.,  {Giuliani} A.,  {Tavani} M.,
  {Molinari} S.,  2019, \aap, 622, A57

\bibitem[\protect\citeauthoryear{{Casse} \& {Paul}}{{Casse} \&
  {Paul}}{1980}]{casse_apj_237_1980}
{Casse} M.,  {Paul} J.~A.,  1980, \apj, 237, 236

\bibitem[\protect\citeauthoryear{{Chevalier} \& {Luo}}{{Chevalier} \&
  {Luo}}{1994}]{chevalier_apj_421_1994}
{Chevalier} R.~A.,  {Luo} D.,  1994, \apj, 421, 225

\bibitem[\protect\citeauthoryear{{Comer\'{o}n} \& {Kaper}}{{Comer\'{o}n} \&
  {Kaper}}{1998}]{comeron_aa_338_1998}
{Comer\'{o}n} F.,  {Kaper} L.,  1998, \aap, 338, 273

\bibitem[\protect\citeauthoryear{{Cox}, {Kerschbaum}, {van Marle}, {Decin},
  {Ladjal}, {Mayer}, {Groenewegen}, {van Eck}, {Royer}, {Ottensamer}, {Ueta},
  {Jorissen}, {Mecina}, {Meliani}, {Luntzer}, {Blommaert}, {Posch},
  {Vandenbussche} \& {Waelkens}}{{Cox} et~al.}{2012}]{2012A&A...537A..35C}
{Cox} N.~L.~J.,  {Kerschbaum} F.,  {van Marle} A.-J.,  {Decin} L.,  {Ladjal}
  D.,  {Mayer} A.,  {Groenewegen} M.~A.~T.,  {van Eck} S.,  {Royer} P.,
  {Ottensamer} R.,  {Ueta} T.,  {Jorissen} A.,  {Mecina} M.,  {Meliani} Z.,
  {Luntzer} A.,  {Blommaert} J.~A.~D.~L.,  {Posch} T.,  {Vandenbussche} B.,
  {Waelkens} C.,  2012, \aap, 537, A35

\bibitem[\protect\citeauthoryear{{Cristofari}, {Blasi} \&
  {Caprioli}}{{Cristofari} et~al.}{2021}]{cristofari_aa_650_2021}
{Cristofari} P.,  {Blasi} P.,    {Caprioli} D.,  2021, \aap, 650, A62

\bibitem[\protect\citeauthoryear{{Das}, {Brose}, {Meyer}, {Pohl}, {Sushch} \&
  {Plotko}}{{Das} et~al.}{2022}]{das_aa_661a_2022}
{Das} S.,  {Brose} R.,  {Meyer} D. M.~A.,  {Pohl} M.,  {Sushch} I.,    {Plotko}
  P.,  2022, \aap, 661, A128

\bibitem[\protect\citeauthoryear{{De Becker}, {del Valle}, {Romero}, {Peri} \&
  {Benaglia}}{{De Becker} et~al.}{2017}]{DeBecker_mnras_471_2017}
{De Becker} M.,  {del Valle} M.~V.,  {Romero} G.~E.,  {Peri} C.~S.,
  {Benaglia} P.,  2017, \mnras, 471, 4452

\bibitem[\protect\citeauthoryear{{del Palacio}, {Bosch-Ramon}, {M{\"u}ller} \&
  {Romero}}{{del Palacio} et~al.}{2018}]{palacio_aa_617_2018}
{del Palacio} S.,  {Bosch-Ramon} V.,  {M{\"u}ller} A.~L.,    {Romero} G.~E.,
  2018, \aap, 617, A13

\bibitem[\protect\citeauthoryear{{del Valle} \& {Pohl}}{{del Valle} \&
  {Pohl}}{2018}]{valle_ApJ_864_2018}
{del Valle} M.~V.,  {Pohl} M.,  2018, \apj, 864, 19

\bibitem[\protect\citeauthoryear{{del Valle} \& {Romero}}{{del Valle} \&
  {Romero}}{2012}]{delvalle_aa_543_2012}
{del Valle} M.~V.,  {Romero} G.~E.,  2012, \aap, 543, A56

\bibitem[\protect\citeauthoryear{{del Valle} \& {Romero}}{{del Valle} \&
  {Romero}}{2014}]{delvalle_aa_563_2014}
{del Valle} M.~V.,  {Romero} G.~E.,  2014, \aap, 563, A96

\bibitem[\protect\citeauthoryear{{del Valle}, {Romero} \& {De Becker}}{{del
  Valle} et~al.}{2013}]{delvalle_aa_550_2013}
{del Valle} M.~V.,  {Romero} G.~E.,    {De Becker} M.,  2013, \aap, 550, A112

\bibitem[\protect\citeauthoryear{{del Valle}, {Romero} \& {Santos-Lima}}{{del
  Valle} et~al.}{2015}]{valle_mnras_448_2015}
{del Valle} M.~V.,  {Romero} G.~E.,    {Santos-Lima} R.,  2015, \mnras, 448,
  207

\bibitem[\protect\citeauthoryear{{Drury} \& {Strong}}{{Drury} \&
  {Strong}}{2015}]{drury_icrc_34_2015}
{Drury} L.,  {Strong} A.,  2015, in 34th International Cosmic Ray Conference
  (ICRC2015) Vol.~34 of International Cosmic Ray Conference, {Cosmic-ray
  diffusive reacceleration: a critical look}.
p.~483

\bibitem[\protect\citeauthoryear{{Dwarkadas}}{{Dwarkadas}}{2005}]{dwarkadas_apj_630_2005}
{Dwarkadas} V.~V.,  2005, \apj, 630, 892

\bibitem[\protect\citeauthoryear{{Dwarkadas}}{{Dwarkadas}}{2007}]{dwarkadas_apj_667_2007}
{Dwarkadas} V.~V.,  2007, \apj, 667, 226

\bibitem[\protect\citeauthoryear{{Ferreira}, {Scherer} \&
  {Potgieter}}{{Ferreira} et~al.}{2008}]{ferreira_adspr_41_2008}
{Ferreira} S.~E.~S.,  {Scherer} K.,    {Potgieter} M.~S.,  2008, Advances in
  Space Research, 41, 351

\bibitem[\protect\citeauthoryear{{Forbrich}, {Menten} \& {Reid}}{{Forbrich}
  et~al.}{2008}]{forbrich_aa_477_2008}
{Forbrich} J.,  {Menten} K.~M.,    {Reid} M.~J.,  2008, \aap, 477, 267

\bibitem[\protect\citeauthoryear{{Forbrich}, {Reid}, {Menten}, {Rivilla},
  {Wolk}, {Rau} \& {Chandler}}{{Forbrich} et~al.}{2017}]{Forbrich_apj_844_2017}
{Forbrich} J.,  {Reid} M.~J.,  {Menten} K.~M.,  {Rivilla} V.~M.,  {Wolk} S.~J.,
   {Rau} U.,    {Chandler} C.~J.,  2017, \apj, 844, 109

\bibitem[\protect\citeauthoryear{{Freyer}, {Hensler} \& {Yorke}}{{Freyer}
  et~al.}{2003}]{freyer_apj_594_2003}
{Freyer} T.,  {Hensler} G.,    {Yorke} H.~W.,  2003, \apj, 594, 888

\bibitem[\protect\citeauthoryear{{Freyer}, {Hensler} \& {Yorke}}{{Freyer}
  et~al.}{2006}]{freyer_apj_638_2006}
{Freyer} T.,  {Hensler} G.,    {Yorke} H.~W.,  2006, \apj, 638, 262

\bibitem[\protect\citeauthoryear{{Garcia-Segura} \& {Mac Low}}{{Garcia-Segura}
  \& {Mac Low}}{1995a}]{garciasegura_aa_455_1995a}
{Garcia-Segura} G.,  {Mac Low} M.-M.,  1995a, \apj, 455, 145

\bibitem[\protect\citeauthoryear{{Garcia-Segura} \& {Mac Low}}{{Garcia-Segura}
  \& {Mac Low}}{1995b}]{garciasegura_aa_455_1995b}
{Garcia-Segura} G.,  {Mac Low} M.-M.,  1995b, \apj, 455, 160

\bibitem[\protect\citeauthoryear{{Garcia-Segura}, {Mac Low} \&
  {Langer}}{{Garcia-Segura} et~al.}{1996}]{garciasegura_1996_aa_305f}
{Garcia-Segura} G.,  {Mac Low} M.-M.,    {Langer} N.,  1996, \aap, 305, 229

\bibitem[\protect\citeauthoryear{{Garc{\'\i}a-Segura}, {Ricker} \&
  {Taam}}{{Garc{\'\i}a-Segura} et~al.}{2018}]{garciasegura_apj_860_2018}
{Garc{\'\i}a-Segura} G.,  {Ricker} P.~M.,    {Taam} R.~E.,  2018, \apj, 860, 19

\bibitem[\protect\citeauthoryear{{Garc{\'\i}a-Segura}, {Taam} \&
  {Ricker}}{{Garc{\'\i}a-Segura} et~al.}{2020}]{garciasegura_apj_893_2020}
{Garc{\'\i}a-Segura} G.,  {Taam} R.~E.,    {Ricker} P.~M.,  2020, \apj, 893,
  150

\bibitem[\protect\citeauthoryear{{Green}, {Mackey}, {Haworth}, {Gvaramadze} \&
  {Duffy}}{{Green} et~al.}{2019}]{green_aa_625_2019}
{Green} S.,  {Mackey} J.,  {Haworth} T.~J.,  {Gvaramadze} V.~V.,    {Duffy} P.,
   2019, \aap, 625, A4

\bibitem[\protect\citeauthoryear{{Groh}, {Meynet}, {Ekstr{\"o}m} \&
  {Georgy}}{{Groh} et~al.}{2014}]{groh_aa564_2014}
{Groh} J.~H.,  {Meynet} G.,  {Ekstr{\"o}m} S.,    {Georgy} C.,  2014, \aap,
  564, A30

\bibitem[\protect\citeauthoryear{{Gull} \& {Sofia}}{{Gull} \&
  {Sofia}}{1979}]{gull_apj_230_1979}
{Gull} T.~R.,  {Sofia} S.,  1979, \apj, 230, 782

\bibitem[\protect\citeauthoryear{{Gvaramadze}, {Menten}, {Kniazev}, {Langer},
  {Mackey}, {Kraus}, {Meyer} \& {Kami{\'n}ski}}{{Gvaramadze}
  et~al.}{2014}]{Gvaramadze_2013}
{Gvaramadze} V.~V.,  {Menten} K.~M.,  {Kniazev} A.~Y.,  {Langer} N.,  {Mackey}
  J.,  {Kraus} A.,  {Meyer} D.~M.-A.,    {Kami{\'n}ski} T.,  2014, \mnras, 437,
  843

\bibitem[\protect\citeauthoryear{{H.~E.~S.~S. Collaboration}}{{H.~E.~S.~S.
  Collaboration}}{2018}]{hess_2018AA_612A_8H}
{H.~E.~S.~S. Collaboration} 2018, \aap, 612, A8

\bibitem[\protect\citeauthoryear{{Hamaguchi}, {Corcoran}, {Pittard}, {Sharma},
  {Takahashi}, {Russell}, {Grefenstette}, {Wik}, {Gull}, {Richardson}, {Madura}
  \& {Moffat}}{{Hamaguchi} et~al.}{2018}]{hamaguchi_natas_2_2018}
{Hamaguchi} K.,  {Corcoran} M.~F.,  {Pittard} J.~M.,  {Sharma} N.,  {Takahashi}
  H.,  {Russell} C. M.~P.,  {Grefenstette} B.~W.,  {Wik} D.~R.,  {Gull} T.~R.,
  {Richardson} N.~D.,  {Madura} T.~I.,    {Moffat} A. F.~J.,  2018, Nature
  Astronomy, 2, 731

\bibitem[\protect\citeauthoryear{Harten, Lax \& van Leer}{Harten
  et~al.}{1983}]{hll_ref}
Harten A.,  Lax P.~D.,    van Leer B.,  1983, SIAM Review, 25, 35

\bibitem[\protect\citeauthoryear{{Herbst}, {Baalmann}, {Bykov}, {Engelbrecht},
  {Ferreira}, {Izmodenov}, {Korolkov}, {Levenfish}, {Linsky}, {Meyer},
  {Scherer} \& {Strauss}}{{Herbst} et~al.}{2022}]{herbst_ssrv_218_2022}
{Herbst} K.,  {Baalmann} L.~R.,  {Bykov} A.,  {Engelbrecht} N.~E.,  {Ferreira}
  S. E.~S.,  {Izmodenov} V.~V.,  {Korolkov} S.~D.,  {Levenfish} K.~P.,
  {Linsky} J.~L.,  {Meyer} D. M.~A.,  {Scherer} K.,    {Strauss} R. D.~T.,
  2022, \ssr, 218, 29

\bibitem[\protect\citeauthoryear{{Herbst}, {Scherer}, {Ferreira}, {Baalmann},
  {Engelbrecht}, {Fichtner}, {Kleimann}, {Strauss}, {Moeketsi} \&
  {Mohamed}}{{Herbst} et~al.}{2020}]{herbst_apj_897_2020}
{Herbst} K.,  {Scherer} K.,  {Ferreira} S. E.~S.,  {Baalmann} L.~R.,
  {Engelbrecht} N.~E.,  {Fichtner} H.,  {Kleimann} J.,  {Strauss} R. D.~T.,
  {Moeketsi} D.~M.,    {Mohamed} S.,  2020, \apjl, 897, L27

\bibitem[\protect\citeauthoryear{{Hubrig}, {Sch{\"o}ller}, {Ilyin},
  {Kharchenko}, {Oskinova}, {Langer}, {Gonz{\'a}lez}, {Kholtygin}, {Briquet} \&
  {Magori Collaboration}}{{Hubrig} et~al.}{2013}]{hubrig_aa_551_2013}
{Hubrig} S.,  {Sch{\"o}ller} M.,  {Ilyin} I.,  {Kharchenko} N.~V.,  {Oskinova}
  L.~M.,  {Langer} N.,  {Gonz{\'a}lez} J.~F.,  {Kholtygin} A.~F.,  {Briquet}
  M.,    {Magori Collaboration} 2013, \aap, 551, A33

\bibitem[\protect\citeauthoryear{{Jaffe}, {Banday}, {Leahy}, {Leach} \&
  {Strong}}{{Jaffe} et~al.}{2011}]{Jaffe_mnras_416_2011}
{Jaffe} T.~R.,  {Banday} A.~J.,  {Leahy} J.~P.,  {Leach} S.,    {Strong} A.~W.,
   2011, \mnras, 416, 1152

\bibitem[\protect\citeauthoryear{{Jokipii}}{{Jokipii}}{1968}]{jokipii_apj_152_1968}
{Jokipii} J.~R.,  1968, \apj, 152, 799

\bibitem[\protect\citeauthoryear{{Jokipii}, {Giacalone} \&
  {K{\'o}ta}}{{Jokipii} et~al.}{2004}]{jokipii_apj_611_2004}
{Jokipii} J.~R.,  {Giacalone} J.,    {K{\'o}ta} J.,  2004, \apjl, 611, L141

\bibitem[\protect\citeauthoryear{{Jokipii} \& {K{\'o}ta}}{{Jokipii} \&
  {K{\'o}ta}}{2014}]{jopikii_apj_794_2014}
{Jokipii} J.~R.,  {K{\'o}ta} J.,  2014, \apjl, 794, L4

\bibitem[\protect\citeauthoryear{{Joubaud}, {Grenier}, {Casandjian},
  {Tolksdorf} \& {Schlickeiser}}{{Joubaud} et~al.}{2020}]{joubaud_aa_635_2020}
{Joubaud} T.,  {Grenier} I.~A.,  {Casandjian} J.~M.,  {Tolksdorf} T.,
  {Schlickeiser} R.,  2020, \aap, 635, A96

\bibitem[\protect\citeauthoryear{{Kobulnicky}, {Chick} \&
  {Povich}}{{Kobulnicky} et~al.}{2018}]{kobulnicky_apj_856_2018}
{Kobulnicky} H.~A.,  {Chick} W.~T.,    {Povich} M.~S.,  2018, \apj, 856, 74

\bibitem[\protect\citeauthoryear{{Kobulnicky}, {Chick}, {Schurhammer},
  {Andrews}, {Povich}, {Munari}, {Olivier}, {Sorber}, {Wernke} \&
  {Dale}}{{Kobulnicky} et~al.}{2016}]{kobulnicky_apjs_227_2016}
{Kobulnicky} H.~A.,  {Chick} W.~T.,  {Schurhammer} D.~P.,  {Andrews} J.~E.,
  {Povich} M.~S.,  {Munari} S.~A.,  {Olivier} G.~M.,  {Sorber} R.~L.,  {Wernke}
  H.~N.,    {Dale} D.~A.,  2016, \apjs, 227, 18

\bibitem[\protect\citeauthoryear{{Kobulnicky}, {Gilbert} \&
  {Kiminki}}{{Kobulnicky} et~al.}{2010}]{kobulnicky_apj_710_2010}
{Kobulnicky} H.~A.,  {Gilbert} I.~J.,    {Kiminki} D.~C.,  2010, \apj, 710, 549

\bibitem[\protect\citeauthoryear{{Kobulnicky}, {Schurhammer}, {Baldwin},
  {Chick}, {Dixon}, {Lee} \& {Povich}}{{Kobulnicky}
  et~al.}{2017}]{kobulnicky_aj_154_2017}
{Kobulnicky} H.~A.,  {Schurhammer} D.~P.,  {Baldwin} D.~J.,  {Chick} W.~T.,
  {Dixon} D.~M.,  {Lee} D.,    {Povich} M.~S.,  2017, \aj, 154, 201

\bibitem[\protect\citeauthoryear{{Krymsky}, {Kuzmin}, {Petukhov} \&
  {Turpanov}}{{Krymsky} et~al.}{1979}]{krymsky_ICRC_1979}
{Krymsky} G.~F.,  {Kuzmin} A.~I.,  {Petukhov} S.~I.,    {Turpanov} A.~A.,
  1979, in International Cosmic Ray Conference Vol.~2 of International Cosmic
  Ray Conference, {Physical Principles of Regular Acceleration Mechanism of
  Charged Particles}.
p.~39

\bibitem[\protect\citeauthoryear{{Lennon}, {Becker}, {Butler}, {Eber}, {Groth},
  {Kunze} \& {Kudritzki}}{{Lennon} et~al.}{1991}]{Lennon_aa_252_1991}
{Lennon} D.~J.,  {Becker} S.~T.,  {Butler} K.,  {Eber} F.,  {Groth} H.~G.,
  {Kunze} D.,    {Kudritzki} R.~P.,  1991, \aap, 252, 498

\bibitem[\protect\citeauthoryear{{Longair}}{{Longair}}{2011}]{longair_2011}
{Longair} M.~S.,  2011, {High Energy Astrophysics}

\bibitem[\protect\citeauthoryear{{Mackey}, {Gvaramadze}, {Mohamed} \&
  {Langer}}{{Mackey} et~al.}{2015}]{mackey_sept_2014}
{Mackey} J.,  {Gvaramadze} V.~V.,  {Mohamed} S.,    {Langer} N.,  2015, \aap,
  573, A10

\bibitem[\protect\citeauthoryear{{Marchili}, {Piano}, {Cardillo}, {Giuliani},
  {Molinari} \& {Tavani}}{{Marchili} et~al.}{2018}]{marchilli_aa_615_2018}
{Marchili} N.,  {Piano} G.,  {Cardillo} M.,  {Giuliani} A.,  {Molinari} S.,
  {Tavani} M.,  2018, \aap, 615, A82

\bibitem[\protect\citeauthoryear{{Massi}, {Forbrich}, {Menten},
  {Torricelli-Ciamponi}, {Neidh{\"o}fer}, {Leurini} \& {Bertoldi}}{{Massi}
  et~al.}{2006}]{massi_aa_453_2006}
{Massi} M.,  {Forbrich} J.,  {Menten} K.~M.,  {Torricelli-Ciamponi} G.,
  {Neidh{\"o}fer} J.,  {Leurini} S.,    {Bertoldi} F.,  2006, \aap, 453, 959

\bibitem[\protect\citeauthoryear{{Meyer}}{{Meyer}}{2021a}]{meyer_mnras_507_2021}
{Meyer} D.~M.~A.,  2021a, \mnras, 507, 4697

\bibitem[\protect\citeauthoryear{{Meyer}}{{Meyer}}{2021b}]{2021arXiv210809273M}
{Meyer} D.~M.~A.,  2021b, arXiv e-prints, p. arXiv:2108.09273

\bibitem[\protect\citeauthoryear{{Meyer}, {Kreplin}, {Kraus}, {Vorobyov},
  {Haemmerle} \& {Eisl{\"o}ffel}}{{Meyer} et~al.}{2019}]{meyer_487_MNRAS_2019}
{Meyer} D.~M.~A.,  {Kreplin} A.,  {Kraus} S.,  {Vorobyov} E.~I.,  {Haemmerle}
  L.,    {Eisl{\"o}ffel} J.,  2019, \mnras, 487, 4473

\bibitem[\protect\citeauthoryear{{Meyer}, {Langer}, {Mackey}, {Vel{\'a}zquez}
  \& {Gusdorf}}{{Meyer} et~al.}{2015}]{meyer_mnras_450_2015}
{Meyer} D.~M.-A.,  {Langer} N.,  {Mackey} J.,  {Vel{\'a}zquez} P.~F.,
  {Gusdorf} A.,  2015, \mnras, 450, 3080

\bibitem[\protect\citeauthoryear{{Meyer}, {Mackey}, {Langer}, {Gvaramadze},
  {Mignone}, {Izzard} \& {Kaper}}{{Meyer} et~al.}{2014}]{meyer_2014bb}
{Meyer} D.~M.-A.,  {Mackey} J.,  {Langer} N.,  {Gvaramadze} V.~V.,  {Mignone}
  A.,  {Izzard} R.~G.,    {Kaper} L.,  2014, \mnras, 444, 2754

\bibitem[\protect\citeauthoryear{{Meyer}, {Mignone}, {Kuiper}, {Raga} \&
  {Kley}}{{Meyer} et~al.}{2017}]{meyer_mnras_464_2017}
{Meyer} D.~M.~A.,  {Mignone} A.,  {Kuiper} R.,  {Raga} A.~C.,    {Kley} W.,
  2017, \mnras, 464, 3229

\bibitem[\protect\citeauthoryear{{Meyer}, {Mignone}, {Petrov}, {Scherer},
  {Vel{\'a}zquez} \& {Boumis}}{{Meyer} et~al.}{2021}]{meyer_mnras_506_2021}
{Meyer} D.~M.~A.,  {Mignone} A.,  {Petrov} M.,  {Scherer} K.,  {Vel{\'a}zquez}
  P.~F.,    {Boumis} P.,  2021, \mnras, 506, 5170

\bibitem[\protect\citeauthoryear{{Meyer}, {Petrov} \& {Pohl}}{{Meyer}
  et~al.}{2020}]{meyer_mnras_493_2020}
{Meyer} D.~M.~A.,  {Petrov} M.,    {Pohl} M.,  2020, \mnras, 493, 3548

\bibitem[\protect\citeauthoryear{{Meyer}, {Pohl}, {Petrov} \&
  {Oskinova}}{{Meyer} et~al.}{2021}]{meyer_mnras_502_2021}
{Meyer} D.~M.~A.,  {Pohl} M.,  {Petrov} M.,    {Oskinova} L.,  2021, \mnras,
  502, 5340

\bibitem[\protect\citeauthoryear{{Meyer}, {van Marle}, {Kuiper} \&
  {Kley}}{{Meyer} et~al.}{2016}]{meyer_obs_2016}
{Meyer} D.~M.-A.,  {van Marle} A.-J.,  {Kuiper} R.,    {Kley} W.,  2016,
  \mnras, 459, 1146

\bibitem[\protect\citeauthoryear{{Meyer}, {Vorobyov}, {Elbakyan},
  {Eisl{\"o}ffel}, {Sobolev} \& {St{\"o}hr}}{{Meyer}
  et~al.}{2021}]{meyer_mnras_500_2021}
{Meyer} D.~M.~A.,  {Vorobyov} E.~I.,  {Elbakyan} V.~G.,  {Eisl{\"o}ffel} J.,
  {Sobolev} A.~M.,    {St{\"o}hr} M.,  2021, \mnras, 500, 4448

\bibitem[\protect\citeauthoryear{{Meyer}, {Vorobyov}, {Elbakyan}, {Kraus},
  {Liu}, {Nayakshin} \& {Sobolev}}{{Meyer} et~al.}{2022}]{meyer_mnras_517_2022}
{Meyer} D.~M.~A.,  {Vorobyov} E.~I.,  {Elbakyan} V.~G.,  {Kraus} S.,  {Liu}
  S.~Y.,  {Nayakshin} S.,    {Sobolev} A.~M.,  2022, \mnras, 517, 4795

\bibitem[\protect\citeauthoryear{{Mignone}, {Bodo}, {Massaglia}, {Matsakos},
  {Tesileanu}, {Zanni} \& {Ferrari}}{{Mignone}
  et~al.}{2007}]{mignone_apj_170_2007}
{Mignone} A.,  {Bodo} G.,  {Massaglia} S.,  {Matsakos} T.,  {Tesileanu} O.,
  {Zanni} C.,    {Ferrari} A.,  2007, \apjs, 170, 228

\bibitem[\protect\citeauthoryear{{Mignone}, {Zanni}, {Tzeferacos}, {van
  Straalen}, {Colella} \& {Bodo}}{{Mignone}
  et~al.}{2012}]{migmone_apjs_198_2012}
{Mignone} A.,  {Zanni} C.,  {Tzeferacos} P.,  {van Straalen} B.,  {Colella} P.,
     {Bodo} G.,  2012, \apjs, 198, 7

\bibitem[\protect\citeauthoryear{{Moore}, {Walter}, {Hester}, {Scowen},
  {Dufour} \& {Buckalew}}{{Moore} et~al.}{2002}]{moore_aj_2002}
{Moore} B.~D.,  {Walter} D.~K.,  {Hester} J.~J.,  {Scowen} P.~A.,  {Dufour}
  R.~J.,    {Buckalew} B.~A.,  2002, \aj, 124, 3313

\bibitem[\protect\citeauthoryear{{Morlino}}{{Morlino}}{2021}]{2021arXiv210801870M}
{Morlino} G.,  2021, arXiv e-prints, p. arXiv:2108.01870

\bibitem[\protect\citeauthoryear{{Morlino}, {Blasi}, {Peretti} \&
  {Cristofari}}{{Morlino} et~al.}{2021}]{morlino_mnras_504_2021}
{Morlino} G.,  {Blasi} P.,  {Peretti} E.,    {Cristofari} P.,  2021, \mnras,
  504, 6096

\bibitem[\protect\citeauthoryear{{Moskalenko}, {Strong}, {Ormes} \&
  {Potgieter}}{{Moskalenko} et~al.}{2002}]{Moskalenko_apj_565_2002}
{Moskalenko} I.~V.,  {Strong} A.~W.,  {Ormes} J.~F.,    {Potgieter} M.~S.,
  2002, \apj, 565, 280

\bibitem[\protect\citeauthoryear{{Moutzouri}, {Mackey}, {Carrasco
  Gonz{\'a}lez}, {Gong}, {Brose}, {Zargaryan}, {Toal{\'a}}, {Menten},
  {Gvaramadze} \& {Rugel}}{{Moutzouri} et~al.}{2022}]{2022arXiv220411913M}
{Moutzouri} M.,  {Mackey} J.,  {Carrasco Gonz{\'a}lez} C.,  {Gong} Y.,  {Brose}
  R.,  {Zargaryan} D.,  {Toal{\'a}} J.~A.,  {Menten} K.~M.,  {Gvaramadze}
  V.~V.,    {Rugel} M.~R.,  2022, arXiv e-prints, p. arXiv:2204.11913

\bibitem[\protect\citeauthoryear{{Parker}}{{Parker}}{1958}]{parker_paj_128_1958}
{Parker} E.~N.,  1958, \apj, 128, 664

\bibitem[\protect\citeauthoryear{{Peri}, {Benaglia}, {Brookes}, {Stevens} \&
  {Isequilla}}{{Peri} et~al.}{2012}]{peri_aa_538_2012}
{Peri} C.~S.,  {Benaglia} P.,  {Brookes} D.~P.,  {Stevens} I.~R.,
  {Isequilla} N.~L.,  2012, \aap, 538, A108

\bibitem[\protect\citeauthoryear{{Peri}, {Benaglia} \& {Isequilla}}{{Peri}
  et~al.}{2015}]{peri_aa_578_2015}
{Peri} C.~S.,  {Benaglia} P.,    {Isequilla} N.~L.,  2015, \aap, 578, A45

\bibitem[\protect\citeauthoryear{{Pillitteri}, {Wolk} \&
  {Megeath}}{{Pillitteri} et~al.}{2016}]{pillitteri_apj_820_2016}
{Pillitteri} I.,  {Wolk} S.~J.,    {Megeath} S.~T.,  2016, \apjl, 820, L28

\bibitem[\protect\citeauthoryear{{Pittard}}{{Pittard}}{2019}]{pittard_mnras_488_2019}
{Pittard} J.~M.,  2019, \mnras, 488, 3376

\bibitem[\protect\citeauthoryear{{Pittard}, {Romero} \& {Vila}}{{Pittard}
  et~al.}{2021}]{pittard_mnras_504_2021}
{Pittard} J.~M.,  {Romero} G.~E.,    {Vila} G.~S.,  2021, \mnras, 504, 4204

\bibitem[\protect\citeauthoryear{{Pittard}, {Wareing} \& {Kupilas}}{{Pittard}
  et~al.}{2021}]{2021arXiv210714673P}
{Pittard} J.~M.,  {Wareing} C.~J.,    {Kupilas} M.~M.,  2021, arXiv e-prints,
  p. arXiv:2107.14673

\bibitem[\protect\citeauthoryear{{Pogorelov}, {Fichtner}, {Czechowski},
  {Lazarian}, {Lembege}, {le Roux}, {Potgieter}, {Scherer}, {Stone}, {Strauss},
  {Wiengarten}, {Wurz}, {Zank} \& {Zhang}}{{Pogorelov}
  et~al.}{2017}]{pogorelov_ssrv_212_2017}
{Pogorelov} N.~V.,  {Fichtner} H.,  {Czechowski} A.,  {Lazarian} A.,  {Lembege}
  B.,  {le Roux} J.~A.,  {Potgieter} M.~S.,  {Scherer} K.,  {Stone} E.~C.,
  {Strauss} R.~D.,  {Wiengarten} T.,  {Wurz} P.,  {Zank} G.~P.,    {Zhang} M.,
  2017, \ssr, 212, 193

\bibitem[\protect\citeauthoryear{Powell}{Powell}{1997}]{Powell1997}
Powell K.~G.,  1997, An Approximate Riemann Solver for Magnetohydrodynamics.
Springer Berlin Heidelberg, Berlin, Heidelberg, pp 570--583

\bibitem[\protect\citeauthoryear{{Prajapati}, {Tej}, {del Palacio}, {Benaglia},
  {CH}, {Vig}, {Mandal} \& {Kanti Ghosh}}{{Prajapati}
  et~al.}{2019}]{prajapati_apj_884_2019}
{Prajapati} P.,  {Tej} A.,  {del Palacio} S.,  {Benaglia} P.,  {CH} I.-C.,
  {Vig} S.,  {Mandal} S.,    {Kanti Ghosh} S.,  2019, \apjl, 884, L49

\bibitem[\protect\citeauthoryear{{Reimer}, {Reimer} \& {Pohl}}{{Reimer}
  et~al.}{2007}]{reimer_apss_309_2007}
{Reimer} A.,  {Reimer} O.,    {Pohl} M.,  2007, \apss, 309, 351

\bibitem[\protect\citeauthoryear{{Rivilla}, {Chandler}, {Sanz-Forcada},
  {Jim{\'e}nez-Serra}, {Forbrich} \& {Mart{\'\i}n-Pintado}}{{Rivilla}
  et~al.}{2015}]{rivilla_apj_808_2015}
{Rivilla} V.~M.,  {Chandler} C.~J.,  {Sanz-Forcada} J.,  {Jim{\'e}nez-Serra}
  I.,  {Forbrich} J.,    {Mart{\'\i}n-Pintado} J.,  2015, \apj, 808, 146

\bibitem[\protect\citeauthoryear{{Rogers} \& {Pittard}}{{Rogers} \&
  {Pittard}}{2013}]{rogers_mnras_431_2013}
{Rogers} H.,  {Pittard} J.~M.,  2013, \mnras, 431, 1337

\bibitem[\protect\citeauthoryear{{Rozyczka} \& {Franco}}{{Rozyczka} \&
  {Franco}}{1996}]{rozyczka_apj_469_1996}
{Rozyczka} M.,  {Franco} J.,  1996, \apjl, 469, L127

\bibitem[\protect\citeauthoryear{{S{\'a}nchez-Ayaso}, {del Valle},
  {Mart{\'\i}}, {Romero} \& {Luque-Escamilla}}{{S{\'a}nchez-Ayaso}
  et~al.}{2018}]{sanchezayaso_apj_861_2018}
{S{\'a}nchez-Ayaso} E.,  {del Valle} M.~V.,  {Mart{\'\i}} J.,  {Romero} G.~E.,
    {Luque-Escamilla} P.~L.,  2018, \apj, 861, 32

\bibitem[\protect\citeauthoryear{{Schatzman}}{{Schatzman}}{1963}]{schatzman_anap_26_1963}
{Schatzman} E.,  1963, Annales d'Astrophysique, 26, 234

\bibitem[\protect\citeauthoryear{{Scherer}, {van der Schyff}, {Bomans},
  {Ferreira}, {Fichtner}, {Kleimann}, {Strauss}, {Weis}, {Wiengarten} \&
  {Wodzinski}}{{Scherer} et~al.}{2015}]{scherer_aa_576_2015}
{Scherer} K.,  {van der Schyff} A.,  {Bomans} D.~J.,  {Ferreira} S.~E.~S.,
  {Fichtner} H.,  {Kleimann} J.,  {Strauss} R.~D.,  {Weis} K.,  {Wiengarten}
  T.,    {Wodzinski} T.,  2015, \aap, 576, A97

\bibitem[\protect\citeauthoryear{{Schulz}, {Ackermann}, {Buehler}, {Mayer} \&
  {Klepser}}{{Schulz} et~al.}{2014}]{schultz_aa_565_2014}
{Schulz} A.,  {Ackermann} M.,  {Buehler} R.,  {Mayer} M.,    {Klepser} S.,
  2014, \aap, 565, A95

\bibitem[\protect\citeauthoryear{{Shklovskii}}{{Shklovskii}}{1954}]{shklovskii_1954}
{Shklovskii} I.~S.,  1954, Akademiia Nauk SSSR Doklady, 94, 417

\bibitem[\protect\citeauthoryear{{Shu}}{{Shu}}{1992}]{shu_pavi_book_1992}
{Shu} F.~H.,  1992, {The physics of astrophysics. Volume II: Gas dynamics.}

\bibitem[\protect\citeauthoryear{{Sushch} \& {Brose}}{{Sushch} \&
  {Brose}}{2023}]{sushch_mnras_521_2023}
{Sushch} I.,  {Brose} R.,  2023, \mnras, 521, 2290

\bibitem[\protect\citeauthoryear{{Sushch}, {Brose} \& {Pohl}}{{Sushch}
  et~al.}{2018}]{sushch_aa_618_2018}
{Sushch} I.,  {Brose} R.,    {Pohl} M.,  2018, \aap, 618, A155

\bibitem[\protect\citeauthoryear{{Tang}}{{Tang}}{2019}]{tang_mnras_482_2019}
{Tang} X.,  2019, \mnras, 482, 3843

\bibitem[\protect\citeauthoryear{{Tang} \& {Chevalier}}{{Tang} \&
  {Chevalier}}{2014}]{tang_apj_784_2014}
{Tang} X.,  {Chevalier} R.~A.,  2014, \apjl, 784, L35

\bibitem[\protect\citeauthoryear{{Telezhinsky}, {Dwarkadas} \&
  {Pohl}}{{Telezhinsky} et~al.}{2012a}]{telezhinsky_aph_35_2012}
{Telezhinsky} I.,  {Dwarkadas} V.~V.,    {Pohl} M.,  2012a, Astroparticle
  Physics, 35, 300

\bibitem[\protect\citeauthoryear{{Telezhinsky}, {Dwarkadas} \&
  {Pohl}}{{Telezhinsky} et~al.}{2012b}]{telezhinsky_aa_541_2012}
{Telezhinsky} I.,  {Dwarkadas} V.~V.,    {Pohl} M.,  2012b, \aap, 541, A153

\bibitem[\protect\citeauthoryear{{Telezhinsky}, {Dwarkadas} \&
  {Pohl}}{{Telezhinsky} et~al.}{2013}]{telezhinsky_aa_552_2013}
{Telezhinsky} I.,  {Dwarkadas} V.~V.,    {Pohl} M.,  2013, \aap, 552, A102

\bibitem[\protect\citeauthoryear{{Thornbury} \& {Drury}}{{Thornbury} \&
  {Drury}}{2014}]{thornbury_mnras_442_2014}
{Thornbury} A.,  {Drury} L.~O.,  2014, \mnras, 442, 3010

\bibitem[\protect\citeauthoryear{{Toal{\'a}}, {Guerrero}, {Todt}, {Sabin},
  {Oskinova}, {Chu}, {Ramos-Larios} \& {G{\'o}mez-Gonz{\'a}lez}}{{Toal{\'a}}
  et~al.}{2020}]{toala_mnras_495_2020}
{Toal{\'a}} J.~A.,  {Guerrero} M.~A.,  {Todt} H.,  {Sabin} L.,  {Oskinova}
  L.~M.,  {Chu} Y.~H.,  {Ramos-Larios} G.,    {G{\'o}mez-Gonz{\'a}lez}
  V.~M.~A.,  2020, \mnras, 495, 3041

\bibitem[\protect\citeauthoryear{{Toal{\'a}}, {Oskinova},
  {Gonz{\'a}lez-Gal{\'a}n}, {Guerrero}, {Ignace} \& {Pohl}}{{Toal{\'a}}
  et~al.}{2016a}]{toala_apj_821}
{Toal{\'a}} J.~A.,  {Oskinova} L.~M.,  {Gonz{\'a}lez-Gal{\'a}n} A.,  {Guerrero}
  M.~A.,  {Ignace} R.,    {Pohl} M.,  2016a, \apj, 821, 79

\bibitem[\protect\citeauthoryear{{Toal{\'a}}, {Oskinova},
  {Gonz{\'a}lez-Gal{\'a}n}, {Guerrero}, {Ignace} \& {Pohl}}{{Toal{\'a}}
  et~al.}{2016b}]{toala_apj_821_2016}
{Toal{\'a}} J.~A.,  {Oskinova} L.~M.,  {Gonz{\'a}lez-Gal{\'a}n} A.,  {Guerrero}
  M.~A.,  {Ignace} R.,    {Pohl} M.,  2016b, \apj, 821, 79

\bibitem[\protect\citeauthoryear{{Toal{\'a}}, {Oskinova} \&
  {Ignace}}{{Toal{\'a}} et~al.}{2017}]{tola_apj_838_2017}
{Toal{\'a}} J.~A.,  {Oskinova} L.~M.,    {Ignace} R.,  2017, \apjl, 838, L19

\bibitem[\protect\citeauthoryear{{Uchiyama}, {Blandford}, {Funk}, {Tajima} \&
  {Tanaka}}{{Uchiyama} et~al.}{2010}]{uchiyama_apj_723_2010}
{Uchiyama} Y.,  {Blandford} R.~D.,  {Funk} S.,  {Tajima} H.,    {Tanaka} T.,
  2010, \apjl, 723, L122

\bibitem[\protect\citeauthoryear{{Vaidya}, {Mignone}, {Bodo}, {Rossi} \&
  {Massaglia}}{{Vaidya} et~al.}{2018}]{vaidya_apj_865_2018}
{Vaidya} B.,  {Mignone} A.,  {Bodo} G.,  {Rossi} P.,    {Massaglia} S.,  2018,
  \apj, 865, 144

\bibitem[\protect\citeauthoryear{{van Marle} \& {Keppens}}{{van Marle} \&
  {Keppens}}{2010}]{vanmarle_comp&fluids_2010pdf}
{van Marle} A.~J.,  {Keppens} R.,  2010, ArXiv e-prints

\bibitem[\protect\citeauthoryear{{van Marle}, {Langer}, {Achterberg} \&
  {Garc{\'{\i}}a-Segura}}{{van Marle} et~al.}{2006}]{vanmarle_aa_460_2006}
{van Marle} A.~J.,  {Langer} N.,  {Achterberg} A.,    {Garc{\'{\i}}a-Segura}
  G.,  2006, \aap, 460, 105

\bibitem[\protect\citeauthoryear{{van Marle}, {Langer} \&
  {Garc{\'{\i}}a-Segura}}{{van Marle} et~al.}{2005}]{vanmarle_aa_444_2005}
{van Marle} A.~J.,  {Langer} N.,    {Garc{\'{\i}}a-Segura} G.,  2005, \aap,
  444, 837

\bibitem[\protect\citeauthoryear{{van Marle}, {Langer} \&
  {Garc{\'{\i}}a-Segura}}{{van Marle} et~al.}{2007}]{vanmarle_aa_469_2007}
{van Marle} A.~J.,  {Langer} N.,    {Garc{\'{\i}}a-Segura} G.,  2007, \aap,
  469, 941

\bibitem[\protect\citeauthoryear{{van Marle}, {Meliani}, {Keppens} \&
  {Decin}}{{van Marle} et~al.}{2011}]{vanmarle_apj_734_2011}
{van Marle} A.~J.,  {Meliani} Z.,  {Keppens} R.,    {Decin} L.,  2011, \apjl,
  734, L26

\bibitem[\protect\citeauthoryear{{van Marle}, {Meliani} \& {Marcowith}}{{van
  Marle} et~al.}{2015}]{vanmarle_584_aa_2015}
{van Marle} A.~J.,  {Meliani} Z.,    {Marcowith} A.,  2015, \aap, 584, A49

\bibitem[\protect\citeauthoryear{{van Veelen}, {Langer}, {Vink},
  {Garc{\'{\i}}a-Segura} \& {van Marle}}{{van Veelen}
  et~al.}{2009}]{vanveelen_aa_50_2009}
{van Veelen} B.,  {Langer} N.,  {Vink} J.,  {Garc{\'{\i}}a-Segura} G.,    {van
  Marle} A.~J.,  2009, \aap, 503, 495

\bibitem[\protect\citeauthoryear{{Vargas-Gonz{\'a}lez}, {Forbrich}, {Dzib} \&
  {Bally}}{{Vargas-Gonz{\'a}lez} et~al.}{2021}]{2021csss.confE.234V}
{Vargas-Gonz{\'a}lez} J.,  {Forbrich} J.,  {Dzib} S.,    {Bally} J.,  2021, in
  The 20.5th Cambridge Workshop on Cool Stars, Stellar Systems, and the Sun
  (CS20.5) Cambridge Workshop on Cool Stars, Stellar Systems, and the Sun, {The
  Orion Radio All-Stars: Radio census and proper motions with the VLA and
  nonthermal variability with ALMA and the VLBA}.
p.~234

\bibitem[\protect\citeauthoryear{{Vel{\'a}zquez}, {Meyer}, {Chiotellis},
  {Cruz-{\'A}lvarez}, {Schneiter}, {Toledo-Roy}, {Reynoso} \&
  {Esquivel}}{{Vel{\'a}zquez} et~al.}{2023}]{velazquez_mnras_519_2023}
{Vel{\'a}zquez} P.~F.,  {Meyer} D.~M.~A.,  {Chiotellis} A.,  {Cruz-{\'A}lvarez}
  A.~E.,  {Schneiter} E.~M.,  {Toledo-Roy} J.~C.,  {Reynoso} E.~M.,
  {Esquivel} A.,  2023, \mnras, 519, 5358

\bibitem[\protect\citeauthoryear{{Villagran}, {G{\'o}mez}, {Vel{\'a}zquez},
  {Meyer}, {Chiotellis}, {Raga}, {Esquivel}, {Toledo-Roy}, {Vargas-Rojas} \&
  {Schneiter}}{{Villagran} et~al.}{2024}]{villagran_mnras_527_2024}
{Villagran} M.~A.,  {G{\'o}mez} D.~O.,  {Vel{\'a}zquez} P.~F.,  {Meyer}
  D.~M.~A.,  {Chiotellis} A.,  {Raga} A.~C.,  {Esquivel} A.,  {Toledo-Roy}
  J.~C.,  {Vargas-Rojas} K.~M.,    {Schneiter} E.~M.,  2024, \mnras, 527, 1601

\bibitem[\protect\citeauthoryear{{Vink}}{{Vink}}{2012}]{vink_aarv_20_2012}
{Vink} J.,  2012, \aapr, 20, 49

\bibitem[\protect\citeauthoryear{{Vink}}{{Vink}}{2020}]{vink_2020}
{Vink} J.,  2020, {Physics and Evolution of Supernova Remnants}

\bibitem[\protect\citeauthoryear{{Voelk} \& {Forman}}{{Voelk} \&
  {Forman}}{1982}]{voelk_apj_253_1982}
{Voelk} H.~J.,  {Forman} M.,  1982, \apj, 253, 188

\bibitem[\protect\citeauthoryear{{Vorobyov}}{{Vorobyov}}{2009}]{vorobyov_apj_704_2009}
{Vorobyov} E.~I.,  2009, \apj, 704, 715

\bibitem[\protect\citeauthoryear{{Vorobyov} \& {Basu}}{{Vorobyov} \&
  {Basu}}{2006}]{voroboyov_apj_650_2006}
{Vorobyov} E.~I.,  {Basu} S.,  2006, \apj, 650, 956

\bibitem[\protect\citeauthoryear{{Vorobyov} \& {Basu}}{{Vorobyov} \&
  {Basu}}{2010}]{vorobyov_apj_719_2010}
{Vorobyov} E.~I.,  {Basu} S.,  2010, \apj, 719, 1896

\bibitem[\protect\citeauthoryear{{Vorobyov}, {Elbakyan}, {Plunkett}, {Dunham},
  {Audard}, {Guedel} \& {Dionatos}}{{Vorobyov}
  et~al.}{2018}]{vorobyov_aa_613_2018}
{Vorobyov} E.~I.,  {Elbakyan} V.~G.,  {Plunkett} A.~L.,  {Dunham} M.~M.,
  {Audard} M.,  {Guedel} M.,    {Dionatos} O.,  2018, \aap, 613, A18

\bibitem[\protect\citeauthoryear{{Wareing}, {Pittard} \& {Falle}}{{Wareing}
  et~al.}{2017}]{wareing_mnras_470_2017}
{Wareing} C.~J.,  {Pittard} J.~M.,    {Falle} S.~A.~E.~G.,  2017, \mnras, 470,
  2283

\bibitem[\protect\citeauthoryear{{Weaver}, {McCray}, {Castor}, {Shapiro} \&
  {Moore}}{{Weaver} et~al.}{1977}]{weaver_apj_218_1977}
{Weaver} R.,  {McCray} R.,  {Castor} J.,  {Shapiro} P.,    {Moore} R.,  1977,
  \apj, 218, 377

\bibitem[\protect\citeauthoryear{{Webb}, {Axford} \& {Forman}}{{Webb}
  et~al.}{1985}]{webb_apj_298_1985}
{Webb} G.~M.,  {Axford} W.~I.,    {Forman} M.~A.,  1985, \apj, 298, 684

\bibitem[\protect\citeauthoryear{{Weiler} \& {Sramek}}{{Weiler} \&
  {Sramek}}{1988}]{weiler_araa_25_1988}
{Weiler} K.~W.,  {Sramek} R.~A.,  1988, \araa, 26, 295

\bibitem[\protect\citeauthoryear{{West}, {Campbell}, {Bhaura}, {Kothes},
  {Safi-Harb}, {Stil}, {Taylor}, {Foster}, {Gaensler}, {George}, {Gibson} \&
  {Ricci}}{{West} et~al.}{2022}]{west_apj_941_2022}
{West} J.~L.,  {Campbell} J.~L.,  {Bhaura} P.,  {Kothes} R.,  {Safi-Harb} S.,
  {Stil} J.~M.,  {Taylor} A.~R.,  {Foster} T.,  {Gaensler} B.~M.,  {George}
  S.~J.,  {Gibson} S.~J.,    {Ricci} R.,  2022, \apj, 941, 6

\bibitem[\protect\citeauthoryear{{Wilhelm}, {Telezhinsky}, {Dwarkadas} \&
  {Pohl}}{{Wilhelm} et~al.}{2020}]{wilhelm_aa_639_2020}
{Wilhelm} A.,  {Telezhinsky} I.,  {Dwarkadas} V.~V.,    {Pohl} M.,  2020, \aap,
  639, A124

\bibitem[\protect\citeauthoryear{{Wilkin}}{{Wilkin}}{1996}]{wilkin_459_apj_1996}
{Wilkin} F.~P.,  1996, \apjl, 459, L31

\bibitem[\protect\citeauthoryear{{Yusef-Zadeh}, {Cornwell}, {Reipurth} \&
  {Roth}}{{Yusef-Zadeh} et~al.}{1990}]{Yusef_apj_343_1990}
{Yusef-Zadeh} F.,  {Cornwell} T.~J.,  {Reipurth} B.,    {Roth} M.,  1990,
  \apjl, 348, L61

\bibitem[\protect\citeauthoryear{{Zhekov} \& {Myasnikov}}{{Zhekov} \&
  {Myasnikov}}{1998}]{zhekov_na_3_1998}
{Zhekov} S.~A.,  {Myasnikov} A.~V.,  1998, \na, 3, 57

\bibitem[\protect\citeauthoryear{{Zirakashvili} \& {Ptuskin}}{{Zirakashvili} \&
  {Ptuskin}}{2018}]{zirakashvili_aph_98_2018}
{Zirakashvili} V.~N.,  {Ptuskin} V.~S.,  2018, Astroparticle Physics, 98, 21

\end{thebibliography}
}


\end{document}